\documentclass[12pt]{article}

\usepackage{amsmath}
\usepackage{amssymb}

\usepackage{graphicx}
\usepackage{scicite}
\usepackage{times}
\usepackage[T1]{fontenc}
\usepackage{deluxetable}
\usepackage{aas_macros}
\usepackage{xspace}
\usepackage{pdfpages}

\newcommand{\nc}[2]{\newcommand{#1}{\ensuremath{#2}\xspace}}
\newcommand{\num}[2]{\newcommand{#1}{{#2}\xspace}}
\usepackage[font=small]{caption}

\nc{\tdur}{\Delta T}

\nc{\tdurCirc}{ \Delta T_{\text{circ}} }
\nc{\ep}{t_0}
\nc{\Per}{P}

\newcommand{\df}{\delta F}

\nc{\teff}{T_{\rm eff} }
\nc{\logg}{\log g}
\nc{\sm}{\sim}
\nc{\Pcad}{ P_{\text{cad}} }
\nc{\NstarEff}{N_{\star,\text{eff}}}

\nc{\neKOISM}{274} 
\nc{\Np}{ n_{\text{pl,cell}} }
\nc{\NpAug}{ n_{\text{pl,aug,cell}} }
\nc{\fcell}{f_{\text{cell}} }

\nc{\fcellBa}{f_{\text{cell,Batalha}}}
\nc{\flogA}{d^{2}\fcell / d\log{P} / d\log{R_P} }
\nc{\NstarAmen}{ n_{\star,\text{amen}} }

\nc{\Rsun}{ R_{\odot} }
\nc{\Kepler}{ \textit{Kepler} }
\nc{\Kp}{ \textit{Kp} }
\nc{\SpecMatch}{\tt SpecMatch}
\nc{\TERRA}{\tt TERRA}

\renewcommand{\Re}{\ensuremath{ R_{\oplus} }\xspace} 
\nc{\Rp}{ R_P }
\nc{\Rstar}{R_\star} 
\nc{\Mstar}{M_\star} 

\nc{\rrat}{\Rp / \Rstar}  
\nc{\rratfrac}{ \frac{\Rp}{\Rstar} } 

\nc{\dfsec}{\delta F_{\text{sec}}}
\nc{\dfpri}{\delta F_{\text{pri}}}
\nc{\teffpri}{\text{T}_{\text{eff,1}}}
\nc{\teffsec}{\text{T}_{\text{eff,2}}}
\nc{\teq}{\text{T}_{\text{eq}} } 

\nc{\ncell}{n_{\text{cell}}}
\nc{\nstar}{n_{\star}}
\nc{\PerMax}{P_{\text{max}}}
\renewcommand{\deg}{\ensuremath{^{\circ}\xspace}}

\nc{\FE}{F_{\oplus}}
\nc{\Fp}{F_{P}}
\nc{\PT}{P_{\text{T}}}
\nc{\Lstar}{L_{\star}} 
\nc{\EtaEarthErr}{5.7^{+1.7}_{-2.2}\%}

\nc{\fBigYearExtrap}{6.4^{+0.5}_{-1.2}\%}
\nc{\fBigYearMeas}{5.0\pm2.1\%}
\nc{\fSmMercExtrap}{6.5^{+0.9}_{-1.7}\%}
\nc{\fSmMercMeas}{5.8\pm1.6\%}
\num{\nPinKp}{98,471}
\num{\nSamp}{42,557}
\num{\nOnSi}{188,329}
\num{\nPinKpTeff}{63,915}
\num{\nPin}{155,046}
\num{\neKOI}{836}
\num{\nPlntMyDV}{650}
\num{\nPlntMyDVCentInfo}{609}
\num{\nPlnt}{603}
\num{\nFPR}{115}
\num{\nFPSE}{44}
\num{\nFPV}{11}
\num{\nFPplnt}{603}
\num{\nFPTTV}{5}
\num{\nCentInfo}{654}
\num{\nFPC}{31}
\num{\nTCE}{2184}
\num{\nFPVD}{27}

\begin{document}


# Prevalence of Earth-size planets orbiting Sun-like stars

Erik A. Petigura[a,b,1], Andrew W. Howard[b], and Geoffrey W. Marcy[a]

[a]Astronomy Department, University of California, Berkeley, CA 94720; and [b]Institute for Astronomy, University of Hawaii at Manoa, Honolulu, HI 96822

**Determining whether Earth-like planets are common or rare looms as a touchstone in the question of life in the universe. We searched for Earth-size planets that cross in front of their host stars by examining the brightness measurements of 42,000 stars from National Aeronautics and Space Administration's *Kepler* mission. We found 603 planets, including 10 that are Earth size ($1-2\ R_\oplus$) and receive comparable levels of stellar energy to that of Earth ($0.25-4\ F_\oplus$). We account for *Kepler*'s imperfect detectability of such planets by injecting synthetic planet–caused dimmings into the *Kepler* brightness measurements and recording the fraction detected. We find that $11 \pm 4\%$ of Sun-like stars harbor an Earth-size planet receiving between one and four times the stellar intensity as Earth. We also find that the occurrence of Earth-size planets is constant with increasing orbital period ($P$), within equal intervals of log$P$ up to ~200 d. Extrapolating, one finds $5.7^{+1.7}_{-2.2}\%$ of Sun-like stars harbor an Earth-size planet with orbital periods of 200–400 d.**

extrasolar planets | astrobiology

The National Aeronautics and Space Administration's (NASA's) *Kepler* mission was launched in 2009 to search for planets that transit (cross in front of) their host stars (1–4). The resulting dimming of the host stars is detectable by measuring their brightness, and *Kepler* monitored the brightness of 150,000 stars every 30 min for 4 y. To date, this exoplanet survey has detected more than 3,000 planet candidates (4).

The most easily detectable planets in the *Kepler* survey are those that are relatively large and orbit close to their host stars, especially those stars having lower intrinsic brightness fluctuations (noise). These large, close-in worlds dominate the list of known exoplanets. However, the *Kepler* brightness measurements can be analyzed and debiased to reveal the diversity of planets, including smaller ones, in our Milky Way Galaxy (5–7). These previous studies showed that small planets approaching Earth size are the most common, but only for planets orbiting close to their host stars. Here, we extend the planet survey to *Kepler*'s most important domain: Earth-size planets orbiting far enough from Sun-like stars to receive a similar intensity of light energy as Earth.

## Planet Survey

We performed an independent search of *Kepler* photometry for transiting planets with the goal of measuring the underlying occurrence distribution of planets as a function of orbital period, $P$, and planet radius, $R_P$. We restricted our survey to a set of Sun-like stars (GK type) that are the most amenable to the detection of Earth-size planets. We define GK-type stars as those with surface temperatures $T_{\rm eff} = 4{,}100–6{,}100$ K and gravities log$g = 4.0–4.9$ (log$g$ is the base 10 logarithm of a star's surface gravity measured in cm s$^{-2}$) (8). Our search for planets was further restricted to the brightest Sun-like stars observed by *Kepler* ($Kp = 10–15$ mag). These 42,557 stars (Best42k) have the lowest photometric noise, making them amenable to the detection of Earth-size planets. When a planet crosses in front of its star, it causes a fractional dimming that is proportional to the fraction of the stellar disk blocked, $\delta F = (R_P/R_\star)^2$, where $R_\star$ is the radius of the star. As viewed by a distant observer, the Earth dims the Sun by ~100 parts per million (ppm) lasting 12 h every 365 d.

We searched for transiting planets in *Kepler* brightness measurements using our custom-built TERRA software package described in previous works (6, 9) and in *SI Appendix*. In brief, TERRA conditions *Kepler* photometry in the time domain, removing outliers, long timescale variability (>10 d), and systematic errors common to a large number of stars. TERRA then searches for transit signals by evaluating the signal-to-noise ratio (SNR) of prospective transits over a finely spaced 3D grid of orbital period, $P$, time of transit, $t_0$, and transit duration, $\Delta T$. This grid-based search extends over the orbital period range of 0.5–400 d.

TERRA produced a list of "threshold crossing events" (TCEs) that meet the key criterion of a photometric dimming SNR ratio SNR > 12. Unfortunately, an unwieldy 16,227 TCEs met this criterion, many of which are inconsistent with the periodic dimming profile from a true transiting planet. Further vetting was performed by automatically assessing which light curves were consistent with theoretical models of transiting planets (10). We also visually inspected each TCE light curve, retaining only those exhibiting a consistent, periodic, box-shaped dimming, and rejecting those caused by single epoch outliers, correlated noise, and other data anomalies. The vetting process was applied homogeneously to all TCEs and is described in further detail in *SI Appendix*.

To assess our vetting accuracy, we evaluated the 235 *Kepler* objects of interest (KOIs) among Best42k stars having $P > 50$ d, which had been found by the *Kepler* Project and identified as planet candidates in the official Exoplanet Archive (exoplanetarchive.ipac.caltech.edu; accessed 19 September 2013). Among them, we found four whose light curves are not consistent with being planets. These four KOIs (364.01, 2,224.02, 2,311.01, and 2,474.01) have long periods and small radii (*SI Appendix*). This exercise suggests that our vetting process is robust and that careful scrutiny of the light curves of small planets in long period orbits is useful to identify false positives.

> **Significance**
>
> A major question is whether planets suitable for biochemistry are common or rare in the universe. Small rocky planets with liquid water enjoy key ingredients for biology. We used the National Aeronautics and Space Administration *Kepler* telescope to survey 42,000 Sun-like stars for periodic dimmings that occur when a planet crosses in front of its host star. We found 603 planets, 10 of which are Earth size and orbit in the habitable zone, where conditions permit surface liquid water. We measured the detectability of these planets by injecting synthetic planet-caused dimmings into *Kepler* brightness measurements. We find that 22% of Sun-like stars harbor Earth-size planets orbiting in their habitable zones. The nearest such planet may be within 12 light-years.

Vetting of our TCEs produced a list of 836 eKOIs, which are analogous to KOIs produced by the *Kepler* Project. Each light curve is consistent with an astrophysical transit but could be due to an eclipsing binary (EB), either in the background or gravitationally bound, instead of a transiting planet. If an EB resides within the software aperture of a *Kepler* target star (within ∼10 arcsec), the dimming of the EB can masquerade as a planet transit when diluted by the bright target star. We rejected as likely EBs any eKOIs with these characteristics: radii larger than 20 $R_\oplus$, observed secondary eclipse, or astrometric motion of the target star in and out of transit (*SI Appendix*). This rejection of EBs left 603 eKOIs in our catalog.

*Kepler* photometry can be used to measure $R_P/R_*$ with high precision, but the extraction of planet radii is compromised by poorly known radii of the host stars (11). To determine $R_*$ and $T_{eff}$, we acquired high-resolution spectra of 274 eKOIs using the HIRES spectrometer on the 10-m Keck I telescope. Notably, we obtained spectra of all 62 eKOIs that have $P > 100$ d. For these stars, the ∼35% errors in $R_*$ were reduced to ∼10% by matching spectra to standards.

To measure planet occurrence, one must not only detect planets but also assess what fraction of planets were missed. Missed planets are of two types: those whose orbital planes are so tilted as to avoid dimming the star and those whose transits were not detected in the photometry by TERRA. Both effects can be quantified to establish a statistical correction factor. The first correction can be computed as the geometrical probability that an orbital plane is viewed edge-on enough (from Earth) that the planet transits the star. This probability is $P_T = R_*/a$, where $a$ is the semimajor axis of the orbit.

The second correction is computed by the injection and recovery of synthetic (mock) planet-caused dimmings into real *Kepler* photometry. We injected 40,000 transit-like synthetic dimmings having randomly selected planetary and orbital properties into the actual photometry of our Best42k star sample, with stars selected at random. We measured survey completeness, $C(P, R_P)$, in small bins of $(P, R_P)$, determining the fraction of injected synthetic planets that were discovered by TERRA (*SI Appendix*). Fig. 1 shows the 603 detected planets and the survey completeness, $C$, color-coded as a function of $P$ and $R_P$.

The survey completeness for small planets is a complicated function of $P$ and $R_P$. It decreases with increasing $P$ and decreasing $R_P$ as expected due to fewer transits and less dimming, respectively. It is dangerous to replace this injection and recovery assessment with noise models to determine $C$. Such models are not sensitive to the absolute normalization of $C$ and only provide relative completeness. Models also may not capture the complexities of a multistage transit-finding pipeline that is challenged by correlated, nonstationary, and non-Gaussian noise. Measuring the occurrence of small planets with long periods requires injection and recovery of synthetic transits to determine the absolute detectability of the small signals buried in noise.

### Planet Occurrence

We define planet occurrence, $f$, to be the fraction of stars having a planet within a specified range of orbital period, size, and perhaps other criteria. We report planet occurrence as a function of planet size and orbital period, $f(P, R_P)$ and as a function of planet size and the stellar light intensity (flux) incident on the planet, $f(F_P, R_P)$.

**Planet Occurrence and Orbital Period.** We computed $f(P, R_P)$ in a 6 × 4 grid of $P$ and $R_P$ shown in Fig. 2. We start by first counting the number of detected planets, $n_{cell}$, in each $P$-$R_P$ cell. Then we computed $f(P, R_P)$ by making statistical corrections for planets missed because of nontransiting orbital inclinations and because of the completeness factor, $C$. The first correction augments each detected transiting planet by $1/P_T = a/R_*$, where $P_T$ is the geometric transit probability, to account for planets missed in inclined orbits. Accounting for the completeness, $C$, the occurrence in a cell is $f(P, R_P) = 1/n_* \sum_i a_i/(R_{*,i} C_i)$, where $n_* = 42{,}557$ stars, and the sum is over all detected planets within that cell. Uncertainties in the statistical corrections for $a/R_*$ and for completeness may cause errors in the final occurrence rates of ∼10%. Such errors will be smaller than the Poisson uncertainties in the occurrence of Earth-size planets in long period orbits.

Fig. 2 shows the occurrence of planets, $f(P, R_P)$, within the $P$-$R_P$ plane. Each cell is color-coded to indicate the final planet occurrence: the fraction of stars having a planet with radius and orbital period corresponding to that cell (after correction for both completeness factors). For example, 7.7 ± 1.3% of Sun-like stars have a planet with periods between 25 and 50 d and sizes between 1 and 2 $R_\oplus$.

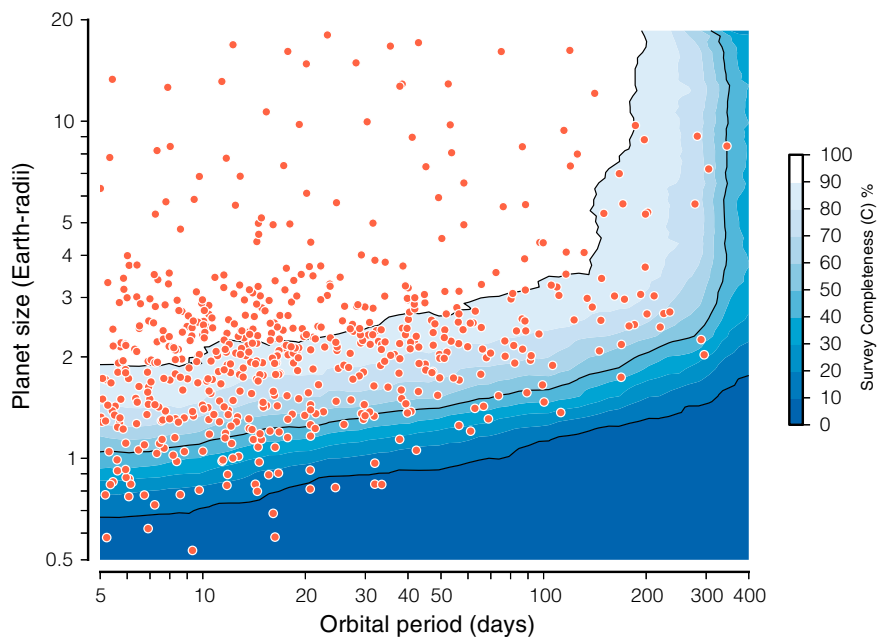

**Fig. 1.** 2D domain of orbital period and planet size, on a logarithmic scale. Red circles show the 603 detected planets in our survey of 42,557 bright Sun-like stars ($Kp$ = 10–15 mag, GK spectral type). The color scale shows survey completeness measured by injection and recovery of synthetic planets into real photometry. Dark regions represent $(P, R_P)$ with low completeness, $C$, where significant corrections for missed planets must be made to compute occurrence. The most common planets detected have orbital $P < 20$ d and $R_P \approx 1 - 3$ $R_\oplus$ (at middle left of graph). However, their detectability is favored by orbital tilt and detection completeness, $C$, that favors detection of such close-in, large planets.

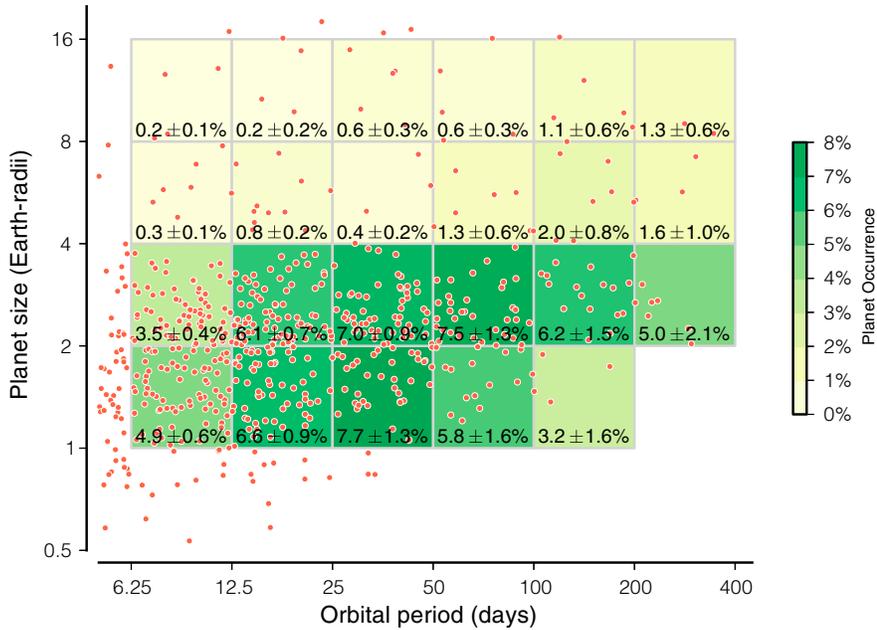

**Fig. 2.** Planet occurrence, $f(P,R_P)$, as a function of orbital period and planet radius for $P = 6.25$–$400$ d and $R_P = 0.5$–$16$ $R_\oplus$. As in Fig. 1, detected planets are shown as red circles. Each cell spans a factor of 2 in orbital period and planet size. Planet occurrence in a cell is given by $f(P,R_P) = 1/n_\star \sum_i a_i / (R_{\star,i} C_i)$, where the sum is over all detected planets within each cell. Here, $a_i/R_i$ is the number of nontransiting planets (for each detected planet) due to large tilt of the orbital plane, $C_i = C(P_i, R_{P,i})$ is the detection completeness factor, and $n_\star = 42{,}557$ stars in the Best42k sample. Cells are colored according to planet occurrence within the cell. We quote planet occurrence within each cell. We do not color cells where the completeness is less than 25%. Among the small planets, 1–2 and 2–4 $R_\oplus$, planet occurrence is constant (within a factor of 2 level) over the entire range of orbital period. This uniformity supports mild extrapolation into the $P = 200$–$400$ d, $R_P = 1$–$2$ $R_\oplus$ domain.
ASTRONOMYWe compute the distribution of planet sizes, including all orbital periods $P = 5$–$100$ d, by summing $f(P,R_P)$ over all periods. The resulting planet size distribution is shown in Fig. 3A. Planets with orbital periods of 5–100 d have a characteristic shape to their size distribution (Fig. 3A). Jupiter-sized planets (11 $R_\oplus$) are rare, but the occurrence of planets rises steadily with decreasing size down to about 2 $R_\oplus$. The distribution is nearly flat (equal numbers of planets per log $R_P$ interval) for 1–2 $R_\oplus$ planets. We find that $26 \pm 3\%$ of Sun-like stars harbor an Earth-size planet ($1 - 2$ $R_\oplus$) with $P = 5$–$100$ d, compared with $1.6 \pm 0.4\%$ occurrence of Jupiter-size planets ($8 - 16$ $R_\oplus$).

We also computed the distribution of orbital periods, including all planet sizes, by summing each period interval of $f(P,R_P)$ over all planet radii. As shown in Fig. 3B, the occurrence of planets larger than Earth rises from $8.9 \pm 0.7\%$ in the $P = 6.25$–$12.5$ d domain to $13.7 \pm 1.2\%$ in the $P = 12.5$–$25$ d interval and is consistent with constant for larger periods. This rise and plateau feature was observed for $\gtrsim 2$ $R_\oplus$ planets in earlier work (5, 12).

Two effects lead to minor corrections to our occurrence estimates. First, some planets in multitransiting systems are missed by TERRA. Second, a small number of eKOIs are false detections. These two effects are small, and they provide corrections to our occurrence statistics with opposite signs. To illustrate their impact, we consider the small and long period ($P > 50$ d) planets that are the focus of this study.

TERRA detects the highest SNR transiting planet per system, so additional transiting planets that cause lower SNR transits are not included in our occurrence measurement. Using the *Kepler* Project catalog (Exoplanet Archive), we counted the number of planets within the same cells in $P$ and $R_P$ as Fig. 2, noting those that did not yield the highest SNR in the system. Inclusion of

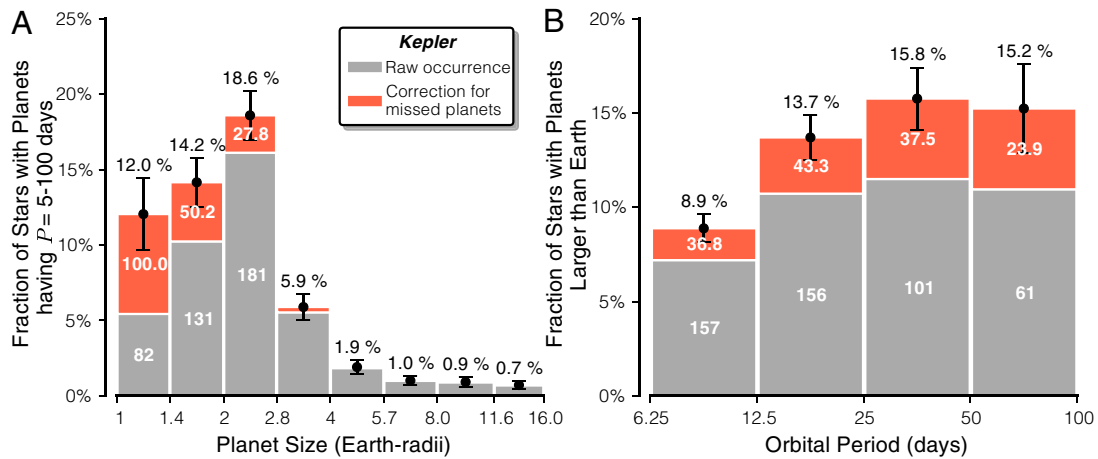

**Fig. 3.** The measured distributions of planet sizes (*A*) and orbital periods (*B*) for $R_P > 1$ $R_\oplus$ and $P = 5$–$100$ d. Heights of the bars represent the fraction of Sun-like stars harboring a planet within a given $P$ or $R_P$ domain. The gray portion of the bars show planet occurrence without correction for survey completeness, i.e., for $C = 1$. The red region shows the correction to account for missed planets, $1/C$. Bars are annotated to reflect the number of planets detected (gray bars) and missed (red bars). The occurrence of planets of different sizes rises by a factor of 10 from Jupiter-size to Earth-sized planets. The occurrence of planets with different orbital periods is constant, within 15%, between 12.5 and 100 d. Due to the small number of detected planets with $R_P = 1 - 2$ $R_\oplus$ and $P > 100$ d (four detected planets), we do not include $P > 100$ d in these marginalized distributions.

footerPetigura et al.   PNAS Early Edition | 3 of 6

these second and third transiting planets boosts the total number of planets per cell (and hence the occurrence) by 21–28% over the $P = 50$–400 d, $R_P = 1$–4 $R_\oplus$ domain (*SI Appendix*).

Even with our careful vetting of eKOIs, the light curves of some false-positive scenarios are indistinguishable from planets. Fressin et al. (7) simulated the contamination of a previous KOI (4) sample by false positives that were not removed by the *Kepler* Project vetting process. They determined that the largest source of false positives for Earth-size planets are physically bound stars with a transiting Neptune-size planet, with an overall false-positive rate of 8.8–12.3%. As we have shown (Fig. 2), the occurrence of Neptune-size planets is nearly constant as a function of orbital period, in $\log P$ intervals. Thus, this false-positive rate is also nearly constant in period. Therefore, we adopt a 10% false-positive rate for planets having $P = 50$–400 d and $R_P = 1$–2 $R_\oplus$. Planet occurrence, shown in Figs. 2 and 3, has not been adjusted to account for false positives or planet multiplicity. The quoted errors reflect only binomial counting uncertainties. Note that for Earth-size planets in the 50–100 and 100–200 d period bins, planet occurrence is 5.8 ± 1.8% and 3.2 ± 1.6%, respectively. Corrections due to false positives or planet multiplicity are smaller than fractional uncertainties due to small number statistics.

**Planet Occurrence and Stellar Light Intensity.** The amount of light energy a planet receives from its host star depends on the luminosity of the star ($L_*$) and the planet-star separation ($a$). Stellar light flux, $F_P$, is given by $F_P = L_*/4\pi a^2$. The intensity of sunlight on Earth is $F_\oplus = 1.36$ kW m$^{-2}$. We compute $L_*$ using $L_* = 4\pi R_*^2 \sigma T_{\text{eff}}^4$, where $\sigma = 5.670 \times 10^{-8}$ Wm$^{-2}$K$^{-4}$ is the Stefan-Boltzmann constant. The dominant uncertainty in $F_P$ is due to $R_*$. Using spectroscopic stellar parameters, we determine $F_P$ to 25% accuracy and to 80% accuracy using photometric parameters. We obtained spectra for all 62 stars hosting planets with $P >$ 100 d, allowing more accurate light intensity measurements.

Fig. 4 shows the 2D domain of stellar light flux incident on our 603 detected planets, along with planet size. The planets in our sample receive a wide range of flux from their host stars, ranging from 0.5 to 700 $F_\oplus$. We highlight the 10 small ($R_P = 1$–2 $R_\oplus$) planets that receive stellar flux comparable to Earth: $F_P = 0.25$–4 $F_\oplus$.

Because only two 1–2 $R_\oplus$ planets have $F_P < 1$ $F_\oplus$, we measure planet occurrence in the domain, 1–2 $R_\oplus$ and 1–4 $F_\oplus$. Correcting for survey completeness, we find that 11 ± 4% of Sun-like stars have a $R_P = 1$–2 $R_\oplus$ planet that receives between one and four times the incident flux as the Earth (*SI Appendix*).

## Interpretation

**Earth-Size Planets with Year-Long Orbital Periods.** Detections of Earth-size planets having orbital periods of $P = 200$–400 d are expected to be rare in this survey. Low survey completeness ($C \approx 10\%$) and low transit probability ($P_T = 0.5\%$) imply that only a few such planets would be expected, even if they are intrinsically common. Indeed, we did not detect any such planets with TERRA, although the radii of three planets (KIC-4478142, KIC-8644545, and KIC-10593626) have $1\sigma$ confidence intervals that extend into the $P = 200$–400 d, $R_P = 1$–2 $R_\oplus$ domain. We can place an upper limit on their occurrence: $f < 12\%$ with 95% confidence using binomial statistics. We would have detected one or two such planets if their occurrence was higher than 12%.

However, one may estimate the occurrence of 1–2 $R_\oplus$ planets with periods of 200–400 d by a modest extrapolation of planet occurrence with $P$. Fig. 5 shows the fraction of stars with 1–2 $R_\oplus$ planets, whose orbital period is less than a maximum period, $P$, on the horizontal axis. This cumulative period distribution shows that 20.4% of Sun-like stars harbor a 1–2 $R_\oplus$ planet with an orbital period, $P < 50$ d. Similarly, 26.2% of Sun-like stars harbor a 1–2 $R_\oplus$ planet with a period less than 100 d. The linear increase in cumulative occurrence implies constant planet occurrence per $\log P$ interval. Extrapolating the cumulative period distribution predicts $5.7^{+1.7}_{-2.2}\%$ occurrence of Earth-size (1–2 $R_\oplus$) planets with orbital periods of ~1 y ($P = 200$–400 d). The details of our extrapolation technique are explained in *SI Appendix*. Extrapolation based on detected planets with $P < 200$ d predicts that $5.7^{+1.7}_{-2.2}\%$ of Sun-like stars have an Earth-size planet on an Earth-like orbit ($P = 200$–400 d).

Naturally, such an extrapolation carries less weight than a direct measurement. However, the loss of *Kepler's* second reaction wheel in May 2013 ended observations shortly after the completion of the nominal 3.5-y mission. We cannot count on any additional *Kepler* data to improve the low completeness to Earth

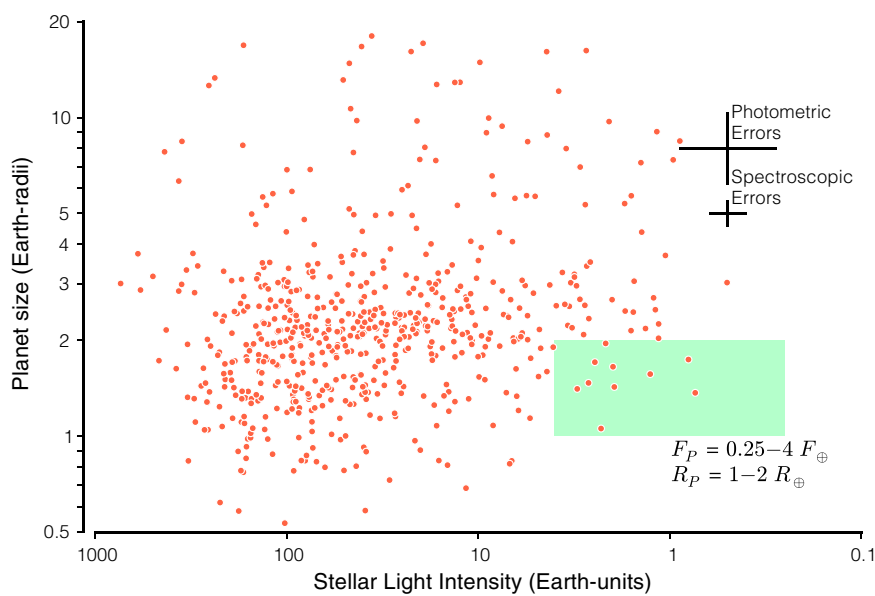

**Fig. 4.** The detected planets (dots) in a 2D domain similar to Figs. 1 and 2. Here, the 2D domain has orbital period replaced by stellar light intensity, incident flux, hitting the planet. The highlighted region shows the 10 Earth-size planets that receive an incident stellar flux comparable to the Earth: flux = 0.25–4.0 times the flux received by the Earth from the Sun. Our uncertainties on stellar flux and planet radii are indicated at the top right.

**Fig. 5.** The fraction of stars having nearly Earth-size planets ($1-2$ $R_\oplus$) with any orbital period up to a maximum period, $P$, on the horizontal axis. Only planets of nearly Earth size ($1-2$ $R_\oplus$) are included. This cumulative distribution reaches 20.2% at $P = 50$ d, meaning 20.4% of Sun-like stars harbor a $1-2$ $R_\oplus$ planet with an orbital period, $P < 50$ d. Similarly, 26.2% of Sun-like stars harbor a $1-2$ $R_\oplus$ planet with a period of $P < 100$ d. The linear increase in this cumulative quantity corresponds to planet occurrence that is constant in equal intervals of $\log P$. One may perform a modest extrapolation into the $P = 200-400$ d range, equivalent to assuming constant occurrence per $\log P$ interval, using all planets with $P > 50$ d. Such an extrapolation predicts that $5.7^{+1.7}_{-2.2}$% of Sun-like stars have a planet with size $1-2$ $R_\oplus$, with an orbital period between $P = 200$ and 400 d.

analog planets beyond what is reported here. Indeed, low survey sensitivity to Earth analogs was the primary reason behind a 4-y extension to the *Kepler* mission. Modest extrapolation is required to understand the prevalence of Earth-size planets with Earth-like orbits around Sun-like stars.

We offer empirical and theoretical justification for extrapolation out to 400 d. As shown in Fig. 2, the prevalence of small planets as a function of $\log P$ is remarkably uniform. To test the reliability of our extrapolation into a region of low completeness, we used the same technique to estimate occurrence in more complete regions of phase space and compared the results with our measured occurrence values.

Again, assuming uniform occurrence per $\log P$ interval, extrapolating the occurrence of $2-4$ $R_\oplus$ planets from 50 to 200 d orbits out to 400 d predicts $6.4^{+0.5}_{-1.2}$% occurrence of planets in the domain of $R_P = 2-4$ $R_\oplus$ and $P = 200-400$ d. The extrapolation is consistent with the measured value of $5.0 \pm 2.1$% to within $1\sigma$ uncertainty. Futhermore, extrapolation based on $1-2$ $R_\oplus$ planets with $P = 12.5-50$ d predicts $6.5^{+0.9}_{-1.7}$% occurrence within the $P = 50-100$ d bin. Again, the measured value of $5.8 \pm 1.6$% agrees to better than $1\sigma$.

Although planet size is governed by nonlinear processes such as runaway gas accretion (13), which favors certain planet sizes over others, no such nonlinear processes occur as a function of orbital period in the range of 200–400 d. Extrapolation out to orbital periods of 200–400 d, although dangerous, seems unlikely to be unrealistic by more than a factor of 2.

**Earth-Size Planets in the Habitable Zone.** Although the details of planetary habitability are debated and depend on planet-specific properties as well as the stochastic nature of planet formation (14), the habitable zone (HZ) is traditionally defined as the set of planetary orbits that permit liquid water on the surface. The precise inner and outer edges of the HZ depend on details of the model (15–18). For solar analog stars, Zsom et al. (17) estimated that the inner edge of the HZ could reside as close as 0.38 astronomical unit (AU) for planets having either a reduced greenhouse effect due to low humidity or a high reflectivity. One AU is the average distance between the Earth and Sun and is

equal to $1.50 \times 10^{11}$ m. Pierrehumbert and Gaidos (18) estimated that the outer edge of the HZ may extend up to 10 AU for planets that are kept warm by efficient greenhouse warming with an H$_2$ atmosphere.

A planet's ability to retain surface liquid water depends, in large part, on the energy received from its host star. We consider a planet to reside in the HZ if it is bathed in a similar level of starlight as Earth. One may adopt $F_P = 0.25-4$ $F_\oplus$ as a simple definition of the HZ, which corresponds to orbital separations of 0.5–2.0 AU for solar analog stars. This definition is more conservative than the range of published HZ boundaries that extend from 0.38 to 10 AU (17, 18). This HZ includes Venus (0.7 AU) and Mars (1.5 AU), which do not currently have surface liquid water. However, Venus may have had liquid water in its past, and there is strong geochemical and geomorphological evidence of liquid water earlier in Mars' history (14).

Previously, we showed that $11 \pm 4$% of stars harbor a planet having an $R_P = 1-2$ $R_\oplus$ and $F_P = 1-4$ $F_\oplus$. Using the definition of $F_P$ and Kepler's third law, $F_P$ is proportional to $P^{-4/3}$. Therefore, uniform occurrence in $\log P$ translates to uniform occurrence in $\log F_P$. We find that the occurrence of $1-2$ $R_\oplus$ planets is constant per $\log F_P$ interval for $F_P = 100-1$ $F_\oplus$. If one was to adopt $F_P = 0.25-4$ $F_\oplus$ as the HZ and extrapolate from the $F_P = 1-4$ $F_\oplus$

**Table 1. Occurrence of small planets in the habitable zone**

| HZ definition | $a_{inner}$ | $a_{outer}$ | $F_{P,inner}$ | $F_{P,outer}$ | $f_{HZ}$ (%) |
|---|---|---|---|---|---|
| Simple | 0.5 | 2 | 4 | 0.25 | 22 |
| Kasting (1993) | 0.95 | 1.37 | 1.11 | 0.53 | 5.8 |
| Kopparapu et al. (2013) | 0.99 | 1.70 | 1.02 | 0.35 | 8.6 |
| Zsom et al. (2013) | 0.38 | | 6.92 | | 26* |
| Pierrehumbert and Gaidos (2011) | | 10 | | 0.01 | ~50[†] |

*Zsom et al. (17) studied the inner edge of the habitable zone. Here we adopt an outer edge of 2 AU from the Simple model.
[†]Pierrehumbert and Gaidos (18) studied the outer edge of the habitable zone. Here we adopt an inner edge of 0.5 AU from the Simple model. Extrapolation out to 10 AU is severely underconstrained. This estimate is highly uncertain and is included for completeness.

domain, then the occurrence of Earth-size planets in the HZ is 22 ± 8% for Sun-like stars.

One may adopt alternative definitions of both the properties of Earth-size planets and the domain of the HZ. We showed previously that the occurrence of planets is approximately constant as a function of both $R_P$ (for $R_P < 2.8\ R_\oplus$) and $P$ (in logarithmic intervals). Thus, the occurrence of planets in this domain is proportional to logarithmic area in the $R_P$–$F_P$ parameter space being considered. For example, the occurrence of planets of size 1.0–1.4 $R_\oplus$ in orbits that receive 0.25–1.0 $F_\oplus$ in stellar flux is 22%/4 = 5.5%. We offer a number of estimates for the prevalence of Earth-size planets in the HZ based on different published definitions of the HZ in Table 1.

Cooler, M dwarf stars also have a high occurrence of Earth-size planets. Based on the *Kepler* planet catalog, Dressing et al. (19) found that $15^{+15}_{-6}\%$ of early M dwarfs have an Earth-size planet ($0.5 - 1.4\ R_\oplus$) in the HZ using a conservative definition of $0.5-1.1\ F_\oplus$ (15) and three times that value when the HZ is expanded to $0.25-1.5\ F_\oplus$ (20). This result is consistent with a Doppler survey that found that $41^{+54}_{-13}\%$ of nearby M dwarfs have planets with masses 1–10 Earth masses ($M_\oplus$) in the HZ (21). Thus, Earth-size planets appear to be common in the HZs of a range of stellar types.

## Conclusions

Using *Kepler* photometry of Sun-like stars (GK-type), we measured the prevalence of planets having different orbital periods and sizes, down to the size of the Earth and out to orbital periods of 1 y. We gathered Keck spectra of all host stars of planets having periods greater than 100 d to accurately determine their radii. The detection of planets with periods longer than 100 d is challenging, and we characterized our sensitivity to such planets by using injection and recovery of synthetic planets in the photometry. After correcting for orbital tilt and detection completeness, we find that 26 ± 3% of Sun-like stars have an Earth-size (1 − 2 $R_\oplus$) planet with $P$ = 5–100 d. We also find that 11 ± 4% of Sun-like stars harbor an Earth-size planet that receives nearly Earth levels of stellar energy ($F_P$ = 1 − 4 $F_\oplus$).

We showed that small planets far outnumber large ones. Only 1.6 ± 0.4% of Sun-like stars harbor a Jupiter-size (8 − 16 $R_\oplus$) planet with $P$ = 5–100 d compared with a 23 ± 3% occurrence of Earth-size planets. This pattern supports the core accretion scenario in which planets form by the accumulation of solids first and gas later in the protoplanetary disk (13, 22–24). The details of this family of models are hotly debated, including the movement of material within the disk, the timescale for planet formation, and the amount of gas accretion in small planets. Our measurement of a constant occurrence of 1–2.8 $R_\oplus$ planets per log$P$ interval establishes an important observational constraint for these models.

The occurrence of Earth-size planets is constant with decreasing stellar light intensity from 100 $F_\oplus$ down to 1 $F_\oplus$. If one was to assume that this pattern continues down to 0.25 $F_\oplus$, then the occurrence of planets having flux levels of 1–0.25 $F_\oplus$ is also 11 ± 4%.

Earth-size planets are common in the *Kepler* field. If the stars in the *Kepler* field are representative of stars in the solar neighborhood, then Earth-size planets are common around nearby Sun-like stars. If one were to adopt a 22% occurrence rate of Earth-size planets in habitable zones of Sun-like stars, then the nearest such planet is expected to orbit a star that is less than 12 light-years from Earth and can be seen by the unaided eye. Future instrumentation to image and take spectra of these Earths need only observe a few dozen nearby stars to detect a sample of Earth-size planets residing in the HZs of their host stars.


 
\pagenumbering{arabic}

\makeatletter 
\renewcommand{\thefigure}{S\@arabic\c@figure}
\makeatother

\renewcommand{\thepage}{S\arabic{page}}  
\renewcommand{\thesection}{S\arabic{section}}  
\baselineskip24pt

\noindent
{\LARGE\textbf{Supporting Information (SI)}}
\vspace{0.5in}

In this supplement to ``Prevalence of Earth-size planets 
orbiting Sun-like stars'' by Petigura et al., we elaborate on the
technical details of our analysis. In Section~\ref{sec:Best42k}, we
define our sample of \nSamp Sun-like stars that are amenable to the
detection of small planets --- the ``Best42k'' stellar sample. In
Section~\ref{sec:TERRA}, we describe the algorithmic components of
\TERRA, our custom pipeline that we used to
find transiting planets within \Kepler
photometry.  Section~\ref{sec:DV} describes ``data validation,'' how
we prune the large number of ``Threshold Crossing Events'' into a list
of \neKOI eKOIs, analogous to KOIs from the \Kepler
Project. Section~\ref{sec:Bogus4} shows four KOIs in the current
online Exoplanet Archive ~\cite{Akeson13} that failed the data
validation step. Section~\ref{sec:AFP} describes the procedure by
which we remove astrophysical false positives from our list of
eKOIs. Section~\ref{sec:RpRefine} describes how we refine our initial
estimate of planet radii using using spectra of the eKOIs coupled
with MCMC-based light curve fitting.  Section~\ref{sec:Completeness}
contains a description of the fundamental component of this study:
measuring the completeness of our planet search by injecting synthetic
transit light curves, caused by planets of all sizes and orbital
periods, directly into the \Kepler photometry, and analyzing the
photometry with our pipeline to determine the fraction of planets
detected. In Section~\ref{sec:Occurrence} we provide details
describing our calculation of planet occurrence and discuss the
effects of multiplanet systems and false positives.

\section{The Best42k Stellar Sample}
\label{sec:Best42k}

We restrict our planet search to Sun-like stars with well-determined
photometric properties and low photometric noise. We select stars
having revised \Kepler Input Catalog (KIC) parameters. Effective temperatures are based on the Pinsonneault et al.~\cite{Pinsonneault12} revisions to the KIC effective temperatures. Surface gravities are based on fits to Yonsei-Yale stellar evolution models~\cite{Yi01} assuming [Fe/H] = $-0.2$. Further details regarding isochrone fitting can be found in Batalha~et~al.~\cite{Batalha12}; Burke~et~al., submitted; and Rowe~et~al., in prep. These revised stellar parameters are tabulated on the Exoplanet Archive with the {\tt prov\_prim} flag set to ``Pinsonneault.'' Out of the \nOnSi stars observed at some point during Q1--Q15, we selected stars that:

\begin{enumerate}
\item Have revised KIC stellar properties. (\nPin stars),
\item \Kp~=~10--15 mag (\nPinKp stars), 
\item \teff~=~4100--6100~K (\nPinKpTeff stars), and
\item \logg~=~4.0--4.9 (cgs) (\nSamp stars).
\end{enumerate}

\begin{figure}
\centering
\includegraphics[width=1\textwidth]{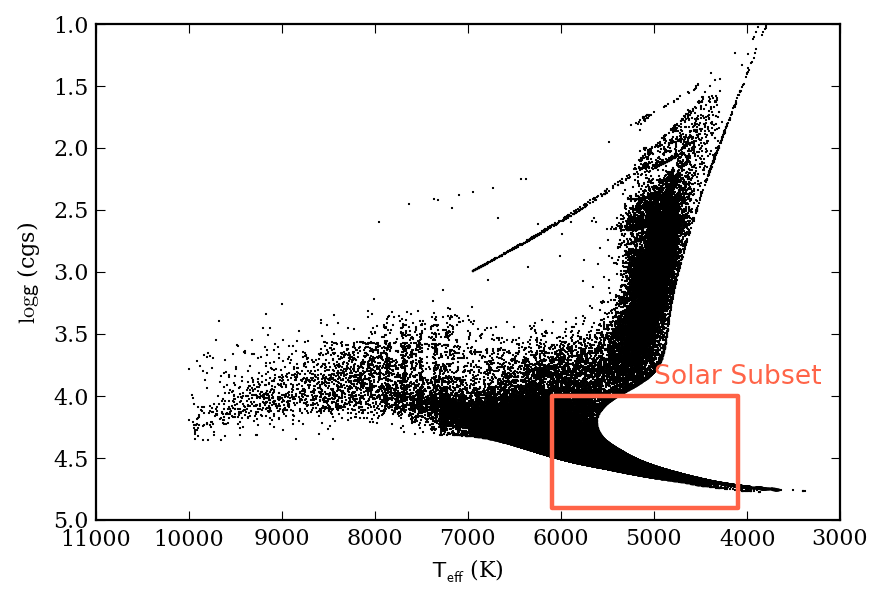}
\caption{Distribution of \nPin stars with revised photometric 
  stellar parameters. The Best42k sample of
  42000 stars is
  made up of Solar-type stars with \teff~=~4100--6100~K,
  \logg~=~4.0--4.9 (cgs), and Kepmag = 10-15 (brighter half of targets).}
\label{fig:SampHR}
\end{figure}

\begin{figure}
\centering
\includegraphics[width=1\textwidth]{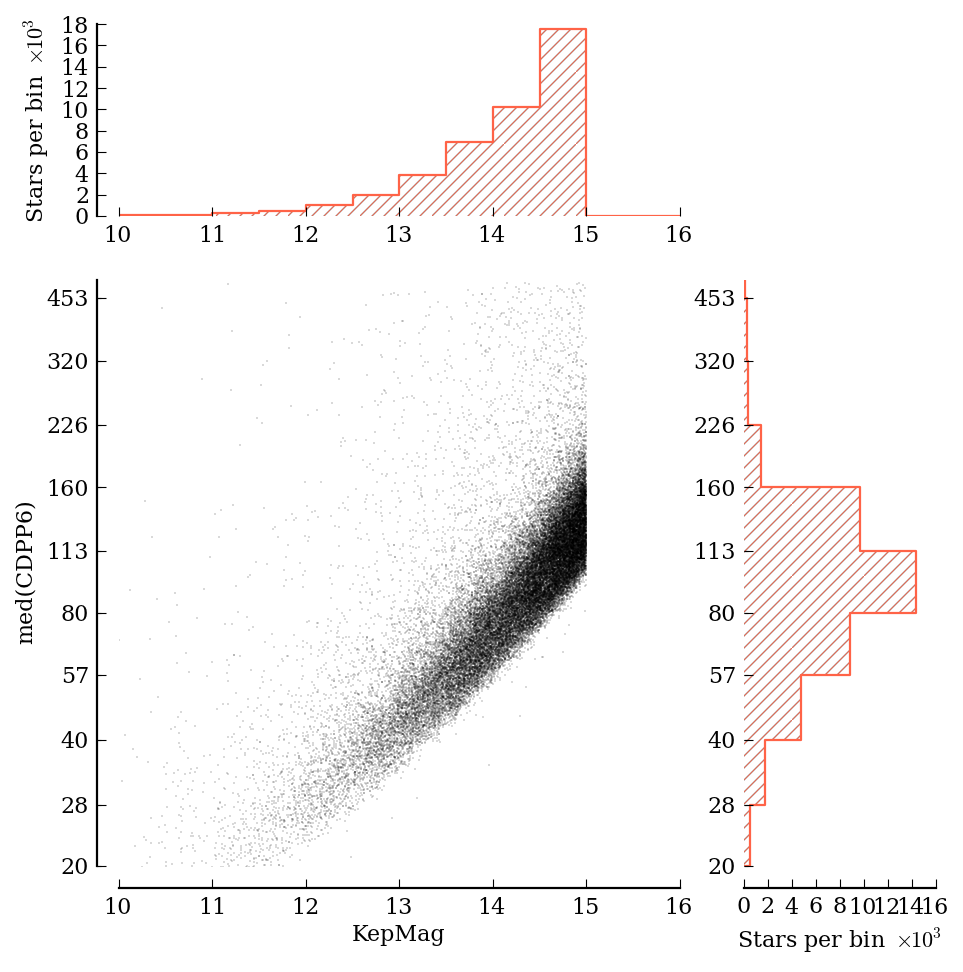}
\caption{Distribution of photometric noise (median quarterly 6 hour CDPP) and brightness \Kp for the \nSamp stars in the Best42k stellar sample. }
\label{fig:SampKpNoise}
\end{figure}

Figure~\ref{fig:SampHR} shows the position of the 155,046 stars with
revised stellar properties along with
the ``solar subset'' corresponding to G and K dwarfs. Figure~\ref{fig:SampKpNoise} shows
the distribution of brightness and noise level of the Best42k stellar
sample.

\clearpage

\section{Planet Search Photometric Pipeline}
\label{sec:TERRA}
We search for planet candidates in the Best42k stellar sample using
the \TERRA pipeline described in detail in Petigura \& Marcy (2012)
and in Petigura, Marcy, and Howard (2013; P13, hereafter)
\cite{Petigura12,Petigura13}. We review the major components of \TERRA
below, noting the changes since P13.

\subsection{Time-domain pre-processing of raw \Kepler Photometry}
\TERRA begins by conditioning the photometry in the
time-domain. \TERRA first searches for single cadence outliers, mostly
due to cosmic rays. \TERRA also searches for abrupt drops in the raw
photometry known as Sudden Pixel Sensitivity Drops (SPSDs) discussed
by Stumpe et al. \cite{Stumpe12}. SPSDs are particularly challenging
since they mimic transit ingress, and aggressive attempts to remove
them run the risk of removing real transits. \TERRA removes the
largest SPSDs, but they remain a source of non-astrophysical false
positives that we remove during manual triage
(Section~\ref{sec:ManualTriage}).

\TERRA also removes trends longer than $\sim$10 days. In P13, this
high-pass filtering was implemented by fitting a spline to the raw
photometry with the knots of the spline separated by 10~days. But in
this work we employ high-pass filtering using Gaussian Process
regression~\cite{Rasmussen06}, which gives finer control over the
timescales removed. We adopt a squared exponential kernel with a 5-day
correlation length. After this high-pass filter, \TERRA identifies
systematic noise modes via principle components analysis on large
number of stars.

\subsection{Grid-based transit search}
We search for periodic box-shaped dimmings by evaluating the
signal-to-noise ratio (SNR) of a putative transit over a finely-spaced
grid of period, \Per; epoch, \ep; and transit duration, \tdur. In P13,
we searched over a period range, \Per~=~5--50~days, and over transit
durations ranging from 1.5--8.8~hr. But in this work, we extend our
search in orbital period to \Per~=~0.5--400~days. Since we search over
nearly three decades in orbital period, and because transit duration
is proportional to $\Per^{1/3}$, we let the range of trial transit
durations vary with period. We break our period range into 10 equal
logarithmic intervals. Then, using photometrically determined
parameters for each star, namely \Mstar and \Rstar, we compute an approximate, expected
transit duration (\tdurCirc) for the simple case of circular orbits
with impact parameter, $b=1$. However, we actually search over \tdur =
0.5--1.5 \tdurCirc to account for a range of impact parameters and
orbital eccentricities and for mis-characterized \Mstar and \Rstar. As
an example, Table~\ref{tab:tdurGrid} shows our trial \tdur for a star
with solar mass and radius.

\begin{deluxetable}{rrr}
\tabletypesize{}
\tablenum{S1}
\tablecaption{\TERRA Grid Search Parameters}
\label{tab:tdurGrid}
\tablewidth{0pt}
\tablehead{
	\colhead{$\Per_1$} &
	\colhead{$\Per_2$} &
	\colhead{Trial Transit Duration (\tdur)}\\
	\colhead{days} &
	\colhead{days} &
	\colhead{long cadence measurements}
 }
\startdata
      5.0 &     7.7 &      [3, 4, 5, 7, 9, 11] \\
      7.7 &    12.0 &         [4, 5, 7, 9, 13] \\
     12.0 &    18.6 &     [4, 5, 7, 9, 13, 15] \\
     18.6 &    28.9 &    [5, 7, 9, 12, 16, 17] \\
     28.9 &    44.7 &   [6, 8, 11, 14, 19, 20] \\
     44.7 &    69.3 &   [7, 9, 12, 16, 22, 23] \\
     69.3 &   107.4 &  [8, 11, 14, 19, 25, 26] \\
    107.4 &   166.5 &  [9, 12, 16, 21, 28, 31] \\
    166.5 &   258.1 & [10, 13, 18, 24, 31, 35] \\
    258.1 &   400.0 & [12, 16, 21, 28, 38, 41] \\
\enddata
\end{deluxetable}
\clearpage

\section{Data Validation}
\label{sec:DV}
If \TERRA detects a (\Per, \ep, \tdur) with SNR > 12, we flag the
light curve for additional scrutiny. While the grid-based component of
\TERRA is well-matched to exoplanet transits, there are other
phenomena that can produce SNR > 12 events and contaminate our planet
sample. We distinguish between two classes of contaminates:
``astrophysical false positives'' such as diluted eclipsing binaries
(EBs), and ``non-astrophysical false positives'' such as noise that
can mimic a transit. We establish a series of quality control measures
called ``Data Validation'' (DV), designed to remove formally strong
dimmings (i.e. SNR > 12) found by the blind photometric pipeline that are
not consistent with an astrophysical transit. DV consists of two
steps:

\begin{enumerate}
\item {\em Machine triage}: Select potential transits by automated cuts.
\item {\em Manual triage}: Manually remove light curves that are inconsistent with a Keplerian transit.
\end{enumerate}

\begin{figure*}
\includegraphics[width=1\textwidth]{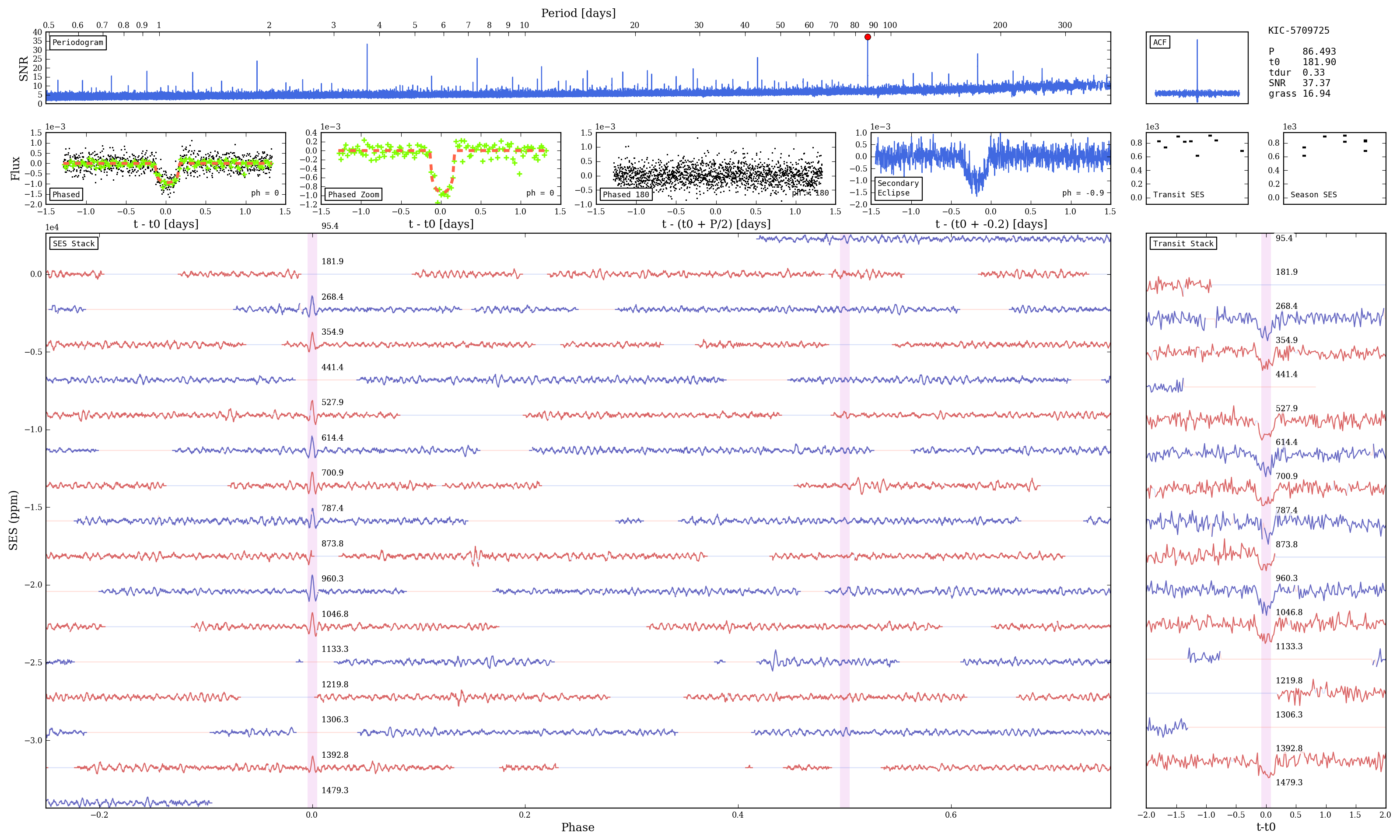}
\caption{DV summary plots for KIC-5709725. Top row: SNR
  ``periodogram'' of box-car photometric search for transiting
  planets, ranging from 0.5 to 400 days. The red dot shows the \Per
  and SNR of the most significant peak in the periodogram at
  \Per~=~86.5~days (also found by the \Kepler Project as
  KOI-555.02). Also visible is a second set of peaks corresponding to
  KOI-555.01 at \Per~=~3.702~days. \TERRA does not search for more than
  one planet per system. Moreover, KOI-555.01 would be excluded from
  our planet sample since \Per~<~5~days. The autocorrelation
  function,``ACF'', plot at upper right shows the circular
  auto-correlation function from the phase folded photometry, used to
  identify secondary eclipses and correlated noise in the
  photometry. Second row: at left ``Phase'' shows the phase--folded
  photometry, where black points represent detrended photometry near
  the time of transit, green symbols show median flux over 30~min
  bins, and the dashed line shows the best-fitting Mandel-Agol transit
  model~\cite{Mandel02}. At second from left, ``Phased Zoom'' shows a
  zoomed y-axis to highlight the transit itself.  For this TCE, the
  transit model is a good match to the photometry. At third from left,
  ``Phased 180'' shows phase folded photometry 180\deg~in orbital
  phase from transit center. At fourth from left, ``Secondary
  eclipse'' shows how \TERRA notches out the putative transit and
  searches for secondary eclipses. We show the photometry folded on
  the second most significant dimming. For KIC-5709725, this phase is
  0.9\deg~relative to the primary transit, so close in phase that the
  primary transit is still visible. This transit does not show signs
  of a secondary eclipse. Transit SES --- transit single event
  statistic as a function of transit number. Conceptually, SES is the
  depth of the transit in ppm, as described by Petigura and Marcy
  \cite{Petigura12}. ``Season SES'' shows the SES statistic grouped
  according to season. Bottom row: At left, ``SES stack'' shows SES
  for the entire light curve, split on the best-fitting transit period
  and stacked so that transit number increases downward. Compelling
  transits appear as a sequence of SES peaks at phase =
  0\deg. ``Transit stack'' shows for TCEs with fewer than 20 transits
  a plot of the \TERRA-calibrated photometry of each transit (transit
  number increases downward).}
\label{fig:eKOIpass}
\end{figure*}

Manual triage is accomplished by inspection of DV summary plots which
contain numerous useful diagnostics necessary to warrant planet
status.  The diagnostics permit a multi-facted evaluation of the
integrity (as a potential planet candidate) of a given dimming
identified by the photometric pipeline. Figure~\ref{fig:eKOIpass}
shows an sample DV report, this for KIC-5709725 that passed
examination.

The product of the DV quality control is a list of ``eKOIs,'' for
which most instrumental events identified preliminarily
and erroneously by the photometric pipeline have been rejected.  The
resulting planet candidates are analogous to the KOIs \Kepler Project.
Astrophysically plausible causes (i.e. transiting planets and
background eclipsing binaries) are retained among our eKOIs.  We
address astrophysical false positives in Section~\ref{sec:AFP}.

\subsection{Machine Triage}
\label{sec:MachineTriage}
\begin{figure*}
\includegraphics[width=1\textwidth]{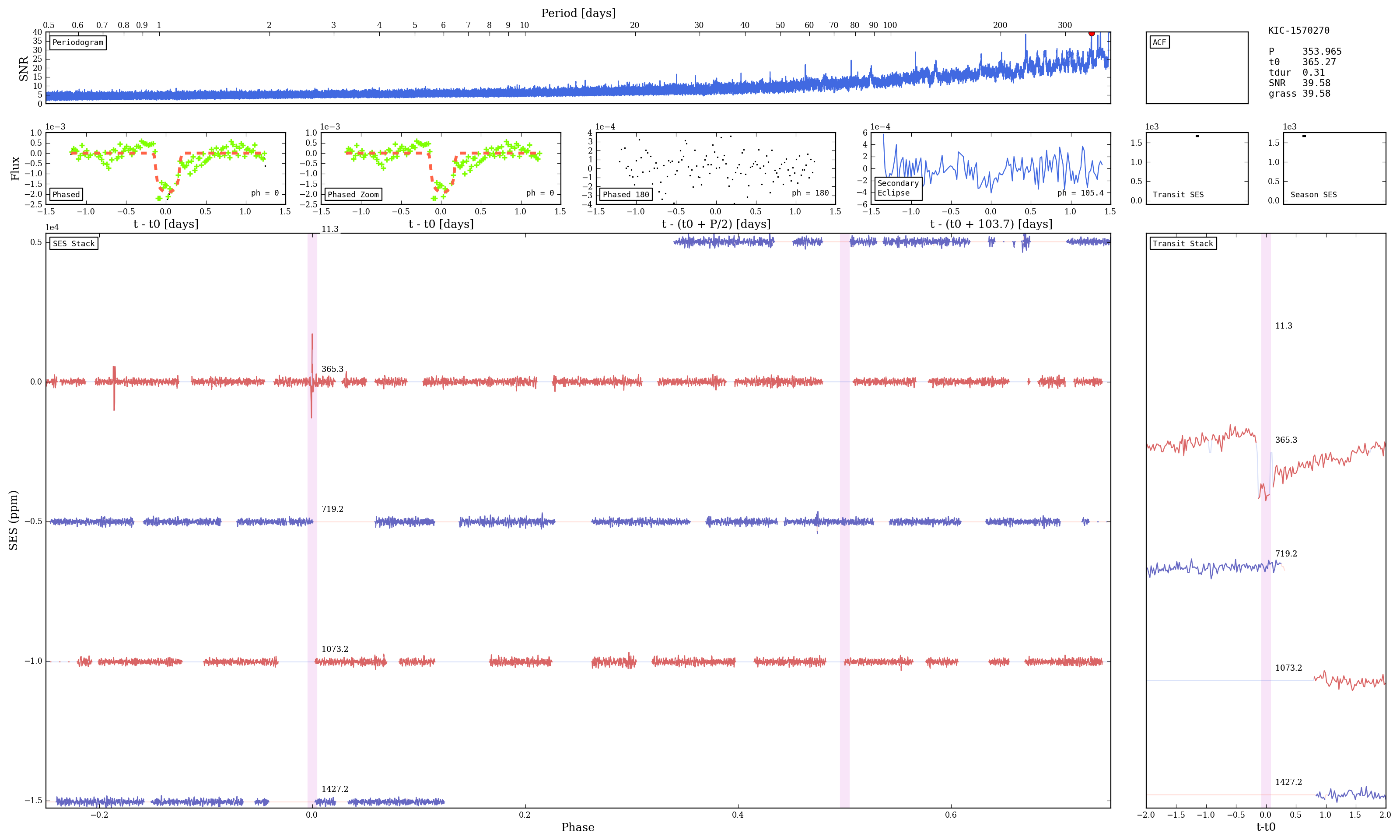}
\caption{DV summary plots (defined in Figure~\ref{fig:eKOIpass}) for
  KIC-1570270 showing a non-astrophysical false positive removed in
  the machine triage step. Here, an uncorrected SPSD at $t = 365.3$~days
  resulted in a SNR$\sim$40 event with \Per~=~353.965 days, seen in
  the SNR periodogram. SES stack plot shows this high SNR TCE is due
  to a single spike in SES due to the SPSD. \Per~=~353.965 days is favored
  over nearby periods because the anomaly aligns with gaps in the
  photometry. We flag cases like this with our {\tt s2n\_on\_grass}
  statistic.  We find several nearby peaks with nearly equal
  SNR. For KIC-1570270, {\tt s2n\_on\_grass} is less than our
  threshold of 1.2 and does not pass machine triage.}
\label{fig:MachineTriage}
\end{figure*}

Prior to the identification of final eKOIs, we carry out machine
triage to identify a set of ``Threshold Crossing Events'' (TCEs) that can be
classified by a human in a reasonable amount of time. TCE status
requires a SNR > 12; however, we find that 16227 light curves (out of
the 42000 target stars) meet this criterion. Outliers and correlated
noise are responsible for the majority of SNR > 12 events. We show set
of diagnostic plots for such an outlier in
Figure~\ref{fig:MachineTriage}. Here, an uncorrected sudden pixel
sensitivity dropoff at $t$~=~365.3~days, raises the noise floor to
SNR$\sim$15 for $\Per \lesssim 100$~days. Its contribution to SNR is
averaged down for shorter periods.

We flag such outliers by comparing the most significant period,
\PerMax, to nearby periods. We call the ratio of the maximum SNR to
the median of the next tallest five peaks between $[\PerMax/1.4 ,
\PerMax \times 1.4$] the {\tt s2n\_on\_grass} statistic. We require
{\tt s2n\_on\_grass} > 1.2 for TCE status. After that cut, 3438 TCEs
remain.  We also require \PerMax > 5 d, which leaves \nTCE TCEs.

\subsection{Manual Triage}
\label{sec:ManualTriage}
The sample of \nTCE TCEs has a significant degree of contamination
from non-astrophysical false positives. In P13, we relied on
aggressive automatic cuts that removed nearly all of the
non-astrophysical false positives (final sample was $\sim 90\%$
pure). However, by comparing our sample to that of Batalha et al. \cite{Batalha12},
we found that these automatic cuts were removing a handful of
compelling planet candidates.

\begin{figure*}
\includegraphics[width=1\textwidth]{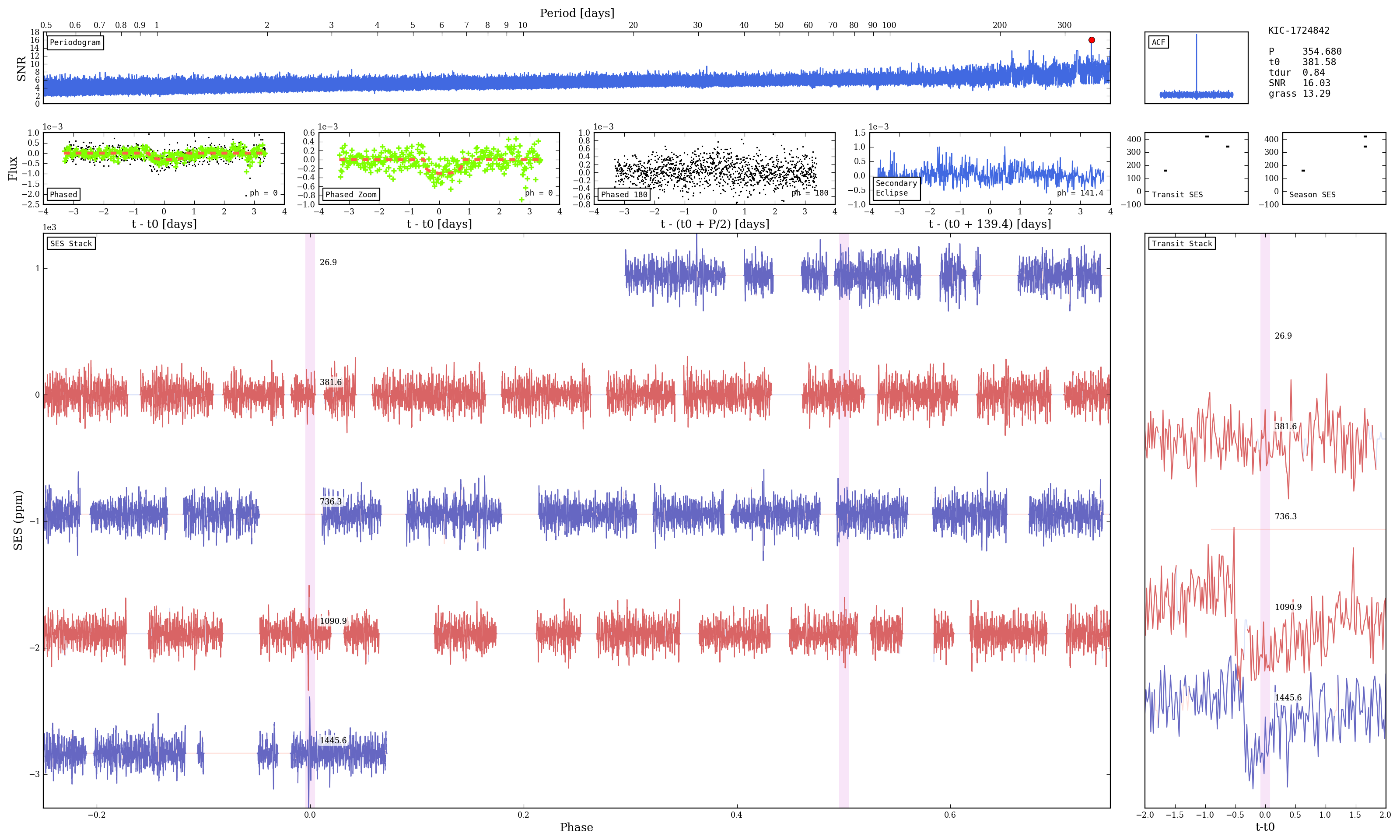}
\caption{DV summary plots (defined in Figure~\ref{fig:eKOIpass}) for
  the KIC-1724842 TCE at \Per~=~354.680~days that we removed during the
  manual triage. This photometry contains two pixel sensitivity drops
  spaced by 354.680~days. These two data anomalies combine to produce
  SNR~=~16.035 event, which is substantially higher than the
  background (``grass'' = 13.294) Since {\tt s2n\_on\_grass} = 1.206 >
  1.2, this event passed our \TERRA software-based triage. However,
  such data anomalies are easily identified by eye.}
\label{fig:ManualTriage}
\end{figure*}

In this work, we aim for higher completeness and rely more heavily on
visual inspection of light curves. We assess whether a TCE is due to a
string of three or more transits or instead caused by outlier(s) such
as SPSDs. Figure~\ref{fig:ManualTriage} shows an example of a light
curve that passed machine triage, but was removed manually. During
manual triage, we do not attempt to distinguish between planets and
astrophysical false positives. The end product is a list of \neKOI
eKOIs, which are analogous to KOIs produced by the \Kepler Project in
that they are highly likely to be astrophysical in origin but false
positives have not been ruled out.

\clearpage

\section{KOIs That Fail Data Validation}
\label{sec:Bogus4}
\begin{figure}
\includegraphics[width=1\textwidth]{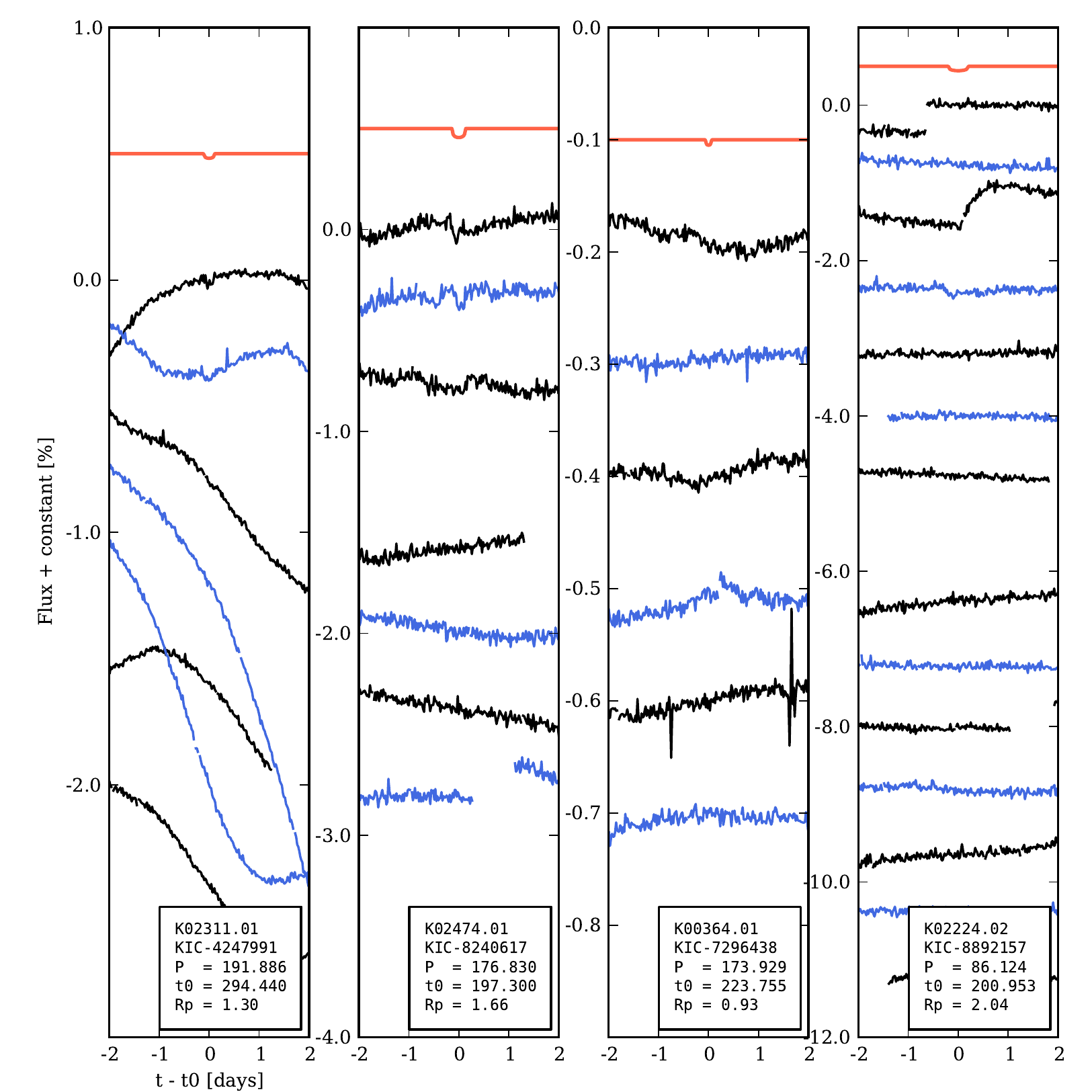}
\caption{Four small KOIs with long orbital periods that fail our
  manual vetting. We show 4-day chunks of raw photometry (``{\tt
    SAP\_FLUX}'' in fits table) around the supposed transits. Transit
  number increases downward. We alternate the use of black and blue
  lines for clarity. The red lines show a Mandel Agol light curve
  model synthesized according to the published transit parameters.}
\label{fig:Bogus4}
\end{figure}

As a cross-check of our DV quality control methods, we performed the
same inspection on 235 KOIs that had been identified by the \Kepler
Project and which appear currently in the online Exoplanet Archive
~\cite{Akeson13}. These KOIs have periods longer than 50 days,
representative of long period transiting planets that enjoy a reduced
number transits (compared to short-period planets) during the 4-year
lifetime of the \Kepler mission.  We found four KOIs, 2311.01,
2474.01, 364.01, and 2224.01, that are not consistent with an
astrophysical transit. We show the raw light curves around the
published ephemerides in Figure~\ref{fig:Bogus4}. All four have
$\Rp~\leq~2.04~\Re$, and three have $\Per~\geq~173$~days. Due to the
small number of KOIs near the habitable zone, inclusion of these KOIs
would bias occurrence measurements upward by a large amount. Vetting
all 3000 KOIs in the Exoplanet Archive requires a substaintial effort,
and is beyond the scope of this paper. However, these four KOIs are a
reminder than detailed, expert, and visual vetting of DV diagnostics
existing KOIs is useful.

\clearpage

\section{Removal of Astrophysical False Positives}
\label{sec:AFP}
\begin{figure*}
\includegraphics[width=1\textwidth]{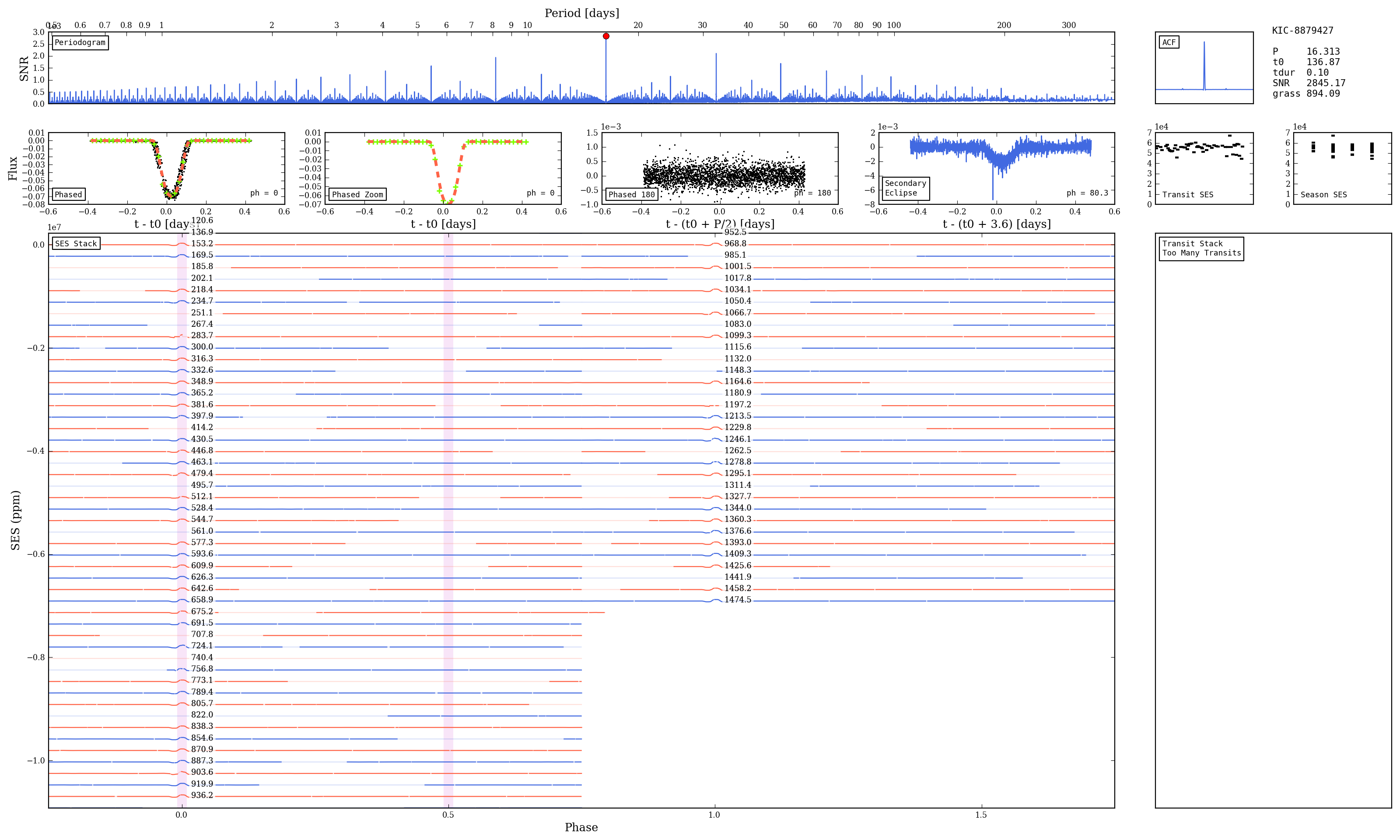}
\caption{DV summary plots (defined in Figure~\ref{fig:eKOIpass}) for
  eKOI KIC-8879427 (\Per = 16.313). The ``Secondary Eclipse'' plot
  shows the second most significant dimming at \Per~=~16.313, offset
  from the primary transit in phase by 80.3\deg. The ratio of the
  primary to secondary eclipses (\dfpri~=~0.07; \dfsec~=~0.002) along
  with the effective temperature of the primary (\teff~=~5995~K) imply
  the transiting object is 2343~K -- too high to be consistent with a
  planet with \Per~=~16.313~days orbital period.}
\label{fig:SecondaryEclipse}
\end{figure*}

\begin{figure*}
\includegraphics[width=1\textwidth]{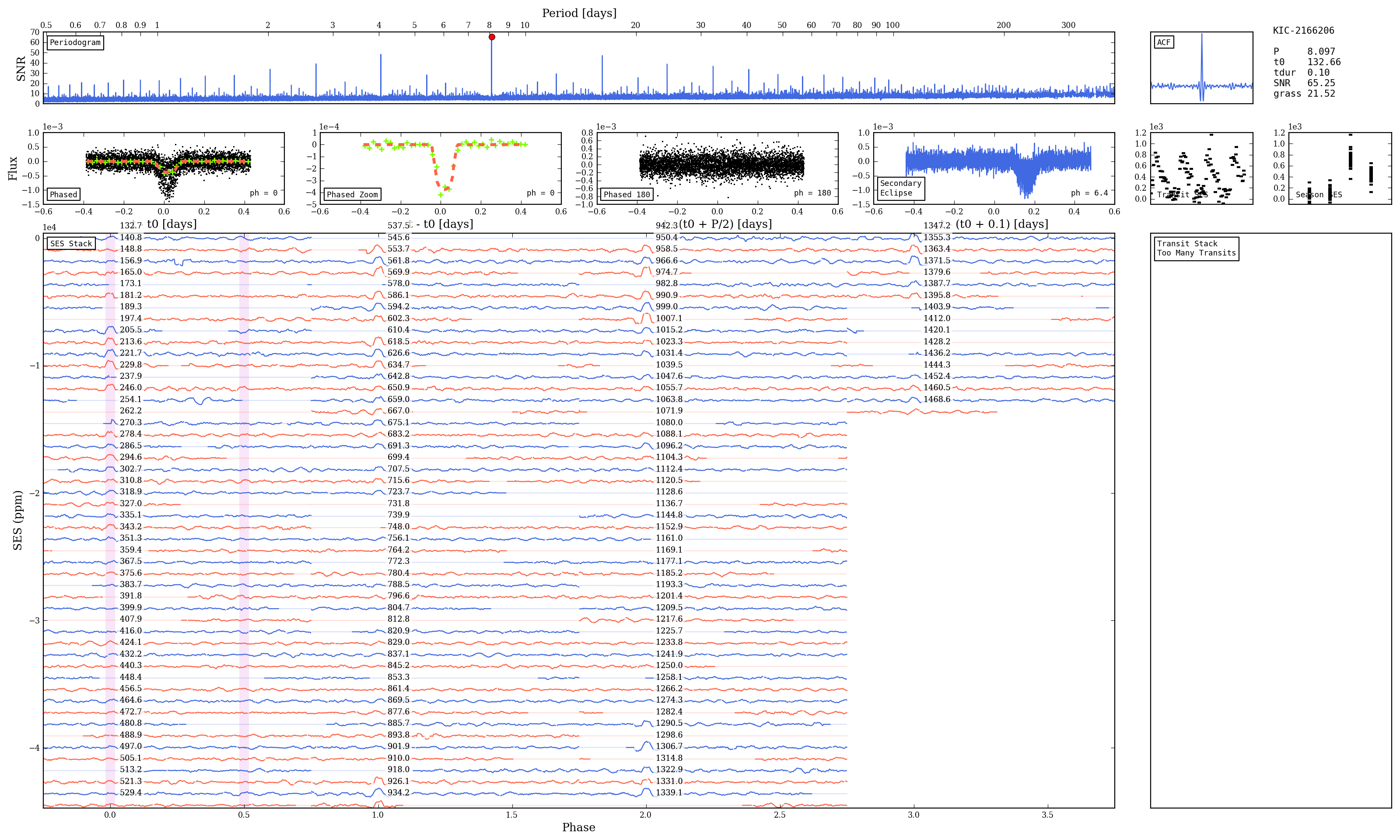}
\caption{DV summary plots (defined in Figure~\ref{fig:eKOIpass}) for
  eKOI KIC-2166206 with season-dependent transit depths. Season SES
  plot shows that the transit depths vary significantly for different
  observing seasons. Since the apparent dimming is a strong function
  of the orientation of the spacecraft, the dimming is likely not
  associated with KIC-2166206, but rather an EB displaced from the
  target by several arc seconds.}
\label{fig:SeasonDependent}
\end{figure*}

We take great care to cleanse our sample of false positives
(FPs). Some transits are so deep ($\df \gtrsim10\%$) that they can
only be caused by an EB. However, if an EB is close enough to a
\Kepler target star, the dimming of the EB can be diluted to the point
where it resembles a planetary transit. For each eKOI, we assess four
indicators of EB status. Here, we list the indicators along with the
number of eKOIs removed from our planet sample due to each cut:
\begin{enumerate}
\item {\em Radius too large (\nFPR)}. We consider any transit where
  the best fit planet radius is larger than 20~\Re to be
  stellar. Planets are generally smaller than
  $1.5~R_{\text{J}}~=~16~\Re$ especially for \Per~>~5~days, where
  planets are less inflated. Our cut at 20 \Re allows some margin of
  safety to account for mis-characterized stellar radii.

\item {\em Secondary eclipse (\nFPSE)}. The expected equilibrium
  temperature for a planet with \Per~>~5~days is too small to produce a
  detectable secondary eclipse. Therefore, the presence of a secondary
  eclipse indicates the eclipsing body is stellar. We search for
  secondary eclipses by masking out the primary transit and searching
  for additional transits at the same period. If an eKOI, such as
  KIC-8879427 shown in Figure~\ref{fig:SecondaryEclipse}, has a
  secondary eclipse, we designate it an EB.

\item {\em Variable depth transits (\nFPVD)}. Since \Kepler
  photometric apertures are typically two or three pixels (8 or 12
  arcsec) on a side, light from neighboring stars can contribute to
  the overall photometry. A faint EB, when diluted with the target
  star's light, can produce a dimming that looks like a planetary
  transit. If the angular separation between the two stars is large
  enough, the EB will contribute a different amount of light at each
  \Kepler orientation. For eKOIs like KIC-2166206 shown in
  Figure~\ref{fig:SeasonDependent}, the contribution of a nearby EB
  results in a season-dependent transit depths. Since the target 
  apertures are defined to include nearly all ($\gtrsim90\%$) of 
  the light from the target star, variations between
  quarters produce a negligible effect  on transits associated with the target star, i.e. fractional changes of $\lesssim 1\%$.

\item {\em Centroid offset (\nFPC)}. \Kepler project DV reports exist
  for nearly all (\nPlntMyDVCentInfo/\nPlntMyDV) of the eKOIs that
  survive the previous cuts and are available on the Exoplanet Archive.  We inspect the
  transit astronomy diagnostics~\cite{Bryson13} for significant motion
  of the transit photocenter in and out of transit. eKOIs with
  significant motion are designated false positives.

\end{enumerate}

We remove a small number of eKOIs (11) with V-shaped transits. Since
planets are so much smaller than their host stars, ingress/egress
durations are short compared to the duration of the transit,
i.e. planetary transits are box-shaped. Stellar eclipses tend to be
V-shaped. Limb-darkening, the 30-minute integration time, and the
possibility of grazing incidence blur this distinction. We assessed
transit shape visually rather than using more detailed approaches
based on light curve fitting and models of Galactic
structure~\cite{Torres11,Morton12}. Only 1.3\% of eKOIs are removed in
this way and are a small effect compared to other uncertainties in our
occurrence measurements.

We also remove five eKOIs with large TTVs. Since \TERRA's light curve
fitting assumes constant period, fits are biased toward smaller planet
radii in the presence of transit timing variations $\gtrsim \tdur$. If
the resulting error is $\gtrsim 25\%$, we remove that eKOI. While
these eKOIs are likely planets, our constant period model results in a
significant bias in derived planet radii. Given the small number of
eKOIs with such large TTVs, our decision to remove them has small
effect on our statistical results based off of hundreds of planets.

\begin{figure}
\centering
\includegraphics[width=1\textwidth]{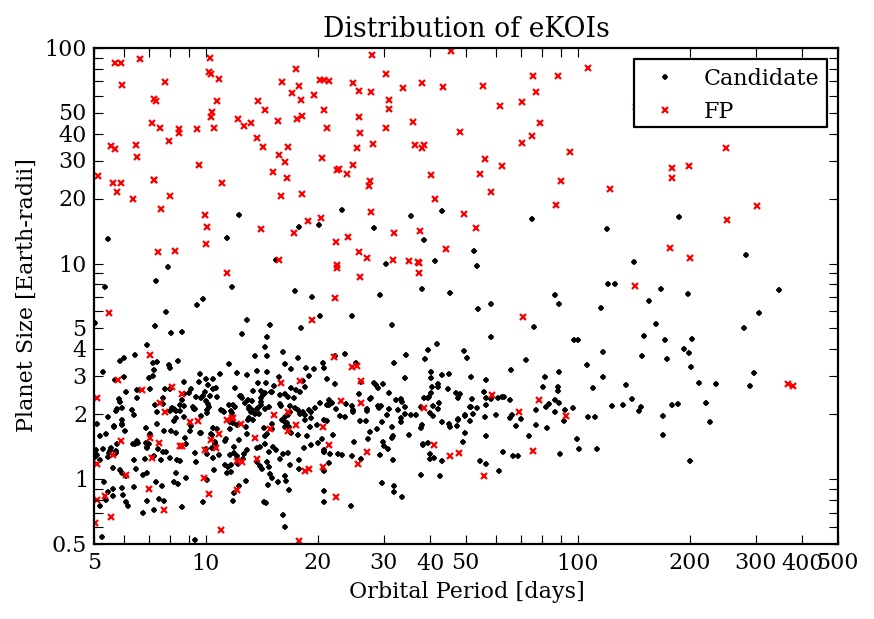}
\caption{Distribution of \TERRA planet candidates as a function of
  planet size and period. Candidates are labeled as black points; FPs
  are labeled as red Xs.}
\label{fig:Catalog}
\end{figure}

We compute planet occurrence from the \nPlnt eKOIs that survive the
above cuts. We show the distribution of \TERRA candidates and FPs on
the \Per--\Rp plane in Figure~\ref{fig:Catalog}. All \neKOI eKOIs are
listed in Table~\ref{tab:eKOI}. For each eKOI, Table~\ref{tab:eKOI}
lists KIC identifier, transit ephemeris, FP designation, Mandel-Agol
fit parameters, adopted host star parameters, and size. We also
crossed checked our eKOIs against the catalog \Kepler team KOIs
accessed from the NASA Exoplanet Archive~\cite{Akeson13} on 13
September 2013. If the \Kepler Project KOI number exists for an
eKOI, we include it in Table~\ref{tab:eKOI}.

\clearpage

\section{Planet Radius Refinement}
\label{sec:RpRefine}
We fit the phase folded transit photometry of each eKOI with a
Mandel-Agol model~\cite{Mandel02}. That model has three free
parameters: \rrat, the planet to star radius radio, $\tau$, the time
for the planet travel a distance \Rstar during during transit; and
$b$, the impact parameter. Following P13, we account for the
covariance among the three parameters using an MCMC exploration of the
parameter posteriors. The error on \rrat in Table~\ref{tab:eKOI}
incorporates the covariance with $\tau$ and $b$.

Because photometry alone only provides the \rrat, knowledge of the
planet population depends heavily on our characterization of their
stellar hosts. We obtained spectra of \neKOISM eKOIs with HIRES on the
Keck I telescope using the standard configuration of the California
Planet Survey (Marcy et al. 2008). These spectra have resolution of
$\sim$50,000 and SNR of $\sim$45/pixel at 5500 \AA. We obtained spectra for
all 62 eKOIs with \Per~>~100~days.

We determine stellar parameters using a routine called {\tt
  \SpecMatch} (Petigura et al., in prep). {\tt \SpecMatch} compares a
target stellar spectrum to a library of $\sim800$ spectra from stars
that span the HR diagram (\teff~=~3500--7500~K; \logg =
2.0--5.0). Parameters for the library stars are determined from LTE
spectral modeling. Once the target spectrum and library spectrum are
placed on the same wavelength scale, we compute $\chi^2$, the sum of
the squares of the pixel-by-pixel differences in normalized
intensity. The weighted mean of the ten spectra with the lowest
$\chi^2$ values is taken as the final value for the effective
temperature, stellar surface gravity, and metallicity. We estimate
{\tt \SpecMatch}-derived stellar radii are uncertain to 10\% RMS,
based on tests of stars having known radii from high resolution
spectroscopy and asteroseismology.

\section{Completeness}
\label{sec:Completeness}

\begin{figure*}
\includegraphics[width=1\textwidth]{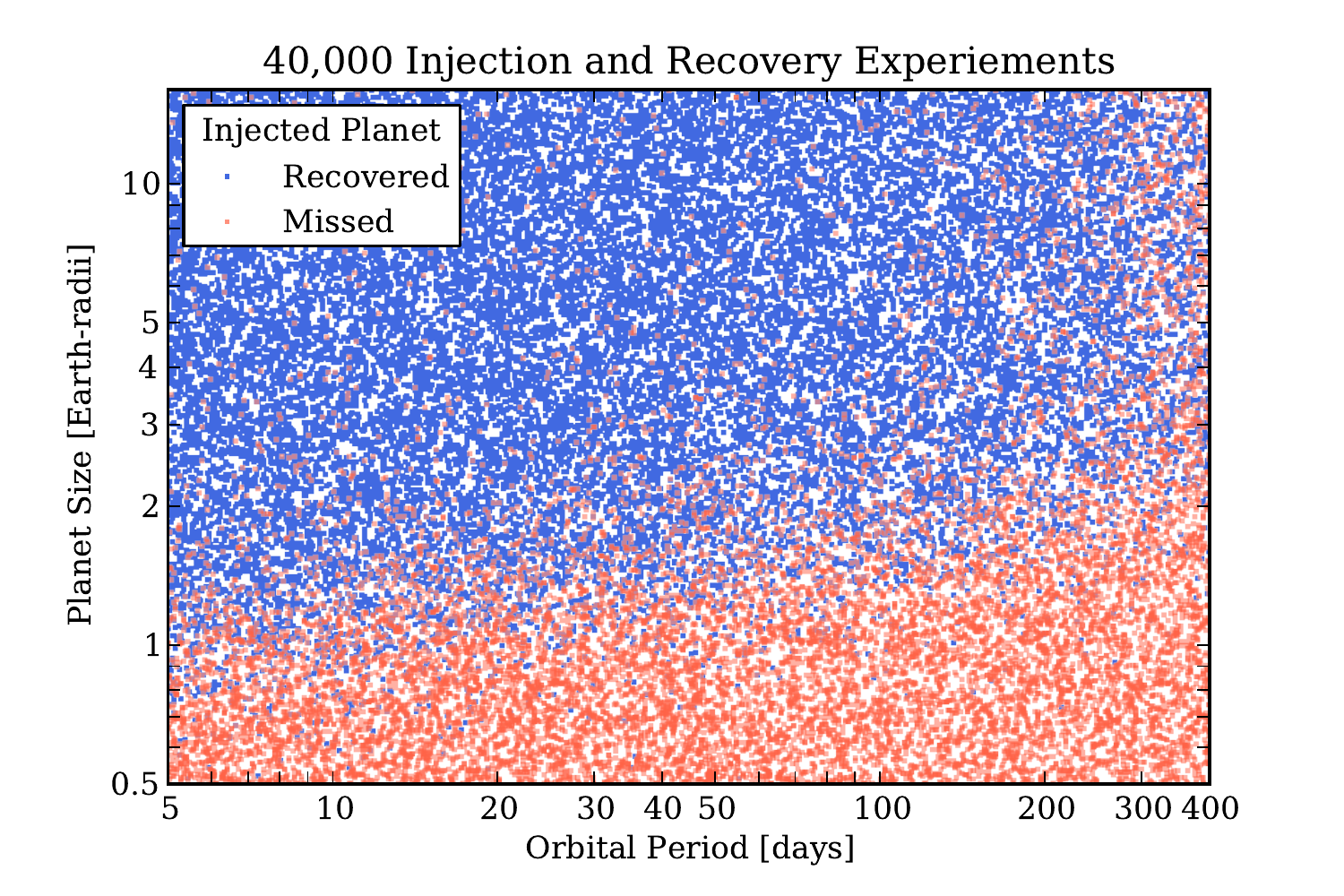}
\caption{\Per and \Rp of 40,000 injected planets color coded by
  whether they were recovered by \TERRA. Completeness over a small
  range in \Per--\Rp is computed by dividing the number of
  successfully recovered transits (blue points) by the total number of
  injected transits (blue and red points). For planets larger than
  2~\Re, completeness is > 50\% out to 400 d. Completeness rapidly
  falls over 1--2~\Re and is $\lesssim$ 10\% for planets smaller than
  1~\Re.}
\label{fig:Completeness}
\end{figure*}

When measuring planet occurrence, understanding the number of missed
planets is as important as the planet catalog itself. We measure
\TERRA's planet finding efficiency as a function of \Per and \Rp using
the injection/recovery framework developed for P13. We briefly review
the key aspects of our pipeline completeness study; for more detail,
please see P13. We generate 40,000 synthetic light curves according to
the following steps:
\begin{enumerate}
\item Select a star randomly from the Best42k sample,
\item draw (\Per,\Rp) randomly from log-uniform distributions over
  5--400 d and 0.5--16 \Re,
\item draw impact parameter and orbital phase randomly from uniform
  distributions over 0--1,
\item synthesize a Mandel-Agol model \cite{Mandel02}, and
\item inject the model into the ``simple aperture photometry'' of a
  random Best42k star.
\end{enumerate}
We process the synthetic photometry with the calibration, grid-based
search, and DV components of \TERRA. We consider a synthetic light
curve successfully recovered if the injected (\Per, \ep) agree with
the recovered (\Per,\ep) to 0.1 days. Figure~\ref{fig:Completeness}
shows the distribution of recovered simulations as a function of
injected planet size and orbital period.

Pipeline completeness is determined in small bins in (\Per,\Rp)-space
by dividing the number of successfully recovered transits by the total
number of injected transits on a bin-by-bin basis. This ratio is
\TERRA's recovery rate of putative planets within the Best42k
sample. Pipeline completeness is higher among a more rarefied sample
of low noise stars. However, a smaller sample of stars yields fewer
planets.

\begin{figure*}
\includegraphics[width=1\textwidth]{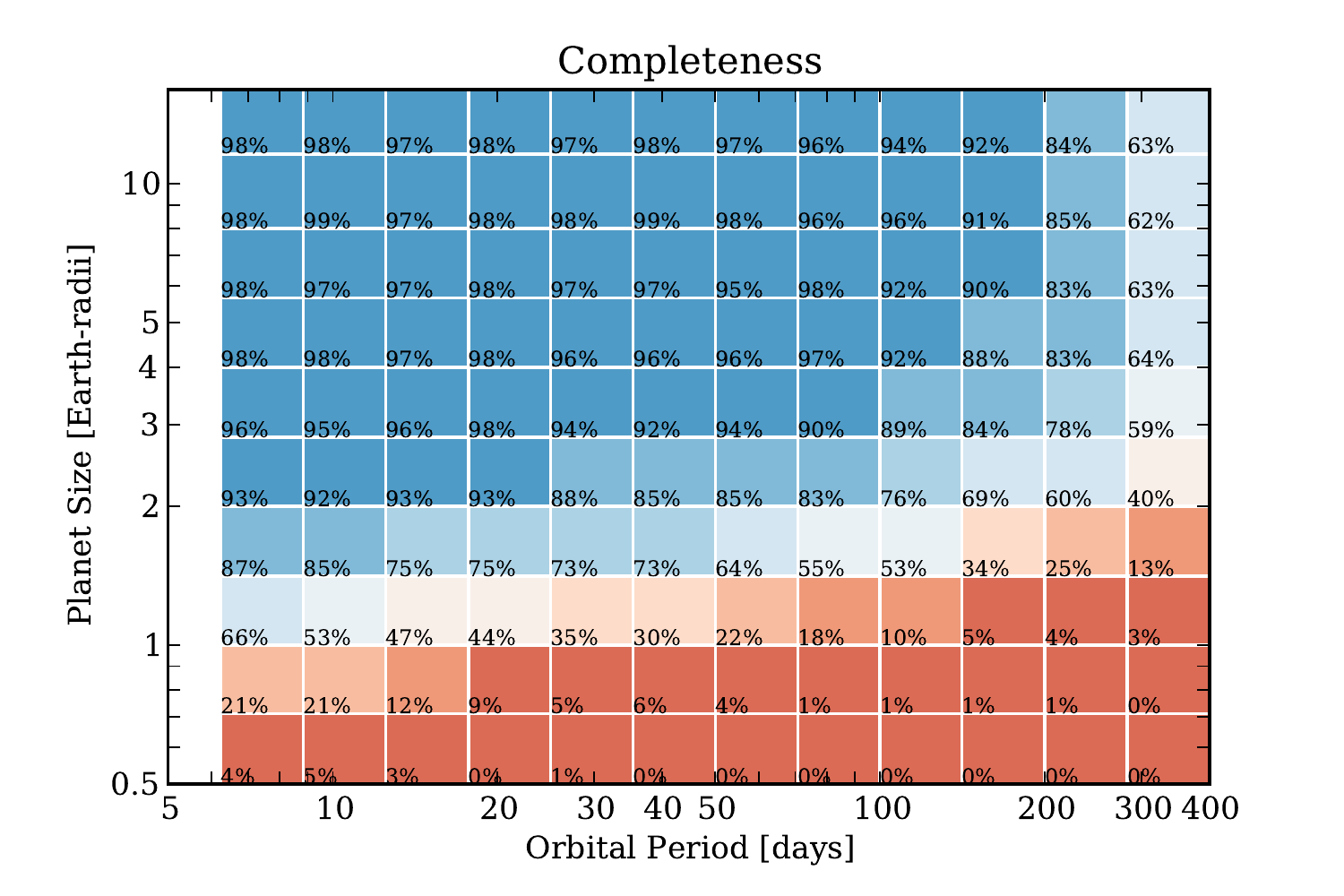}
\caption{Completeness computed over small bins in \Per and \Rp.}
\label{fig:CompletenessBinned}
\end{figure*}

We show survey completeness for a dense grid of \Per and \Rp cells in
Figure~\ref{fig:CompletenessBinned}. Completeness falls toward smaller
\Rp and longer \Per. Above 2 \Re, completeness is greater than 50\%
even for the longest periods searched (except for the \Rp~=~2--2.8
\Re, \Per~=~283--400~days bin). Completeness falls precipitously toward
smaller planet sizes; very few simulated planets smaller than Earth
are recovered. Compared to a 1 \Re planet, a 2 \Re planet produces a
transit with 4 times the SNR and is much easier to detect. For planets
larger than 2 \Re, we note a gradual drop in completeness toward
longer periods, that steepens at $\sim$300 days. Above $\sim$300 days,
the probability that a two or more transits land in data gaps becomes
appreciable, and the completeness falls off more rapidly.

Measuring completeness by injection and recovery captures the vagaries
in planet search pipeline. Real and synthetic transits are treated the
same way, up until the manual triage section. Recall from
Section~\ref{sec:ManualTriage} that \neKOI of \nTCE TCEs pass machine
triage. We perform no such manual inspection of TCEs from the
injection and recovery simulations. A potential concern is that a
planet may pass machine triage, but is accidentally thrown out in
manual triage. Such a planet would be missing from our planet catalog,
but not properly accounted in the occurrence measurement by lower
completeness. However, because our SNR > 12 threshold for TCE status
is high, distinguishing non-astrophysical false positives and eKOIs is
easy. Therefore, we consider it unlikely that they are cut during the
manual triage stage, and do not expect the lack of manual vetting of
the injected TCEs to bias our completeness measurements.

\clearpage

\section{Planet Occurrence}
\label{sec:Occurrence}
Here, we expand on the key planet occurrence results presented in the
main text. We describe our method for extrapolation into the
\Rp~=~1--2~\Re, \Per~=~200--400 day domain. We give additional details
regarding our measurement of the prevalence of Earth-size planets in
the HZ. We also discuss two minor corrections to our occurrence
measurements due to planets in multiplanet systems and false positives
(FPs).

\subsection{Occurrence of Earth-Size Planets on Year-Long Orbits}
In the main text, we reported \EtaEarthErr occurrence of planets with
\Rp~=~1--2~\Re and \Per~=~200--400~days based on extrapolation from
shorter periods. The use of such extrapolation is supported by uniform
planet occurrence per $\log \Per$ interval. Cumulative Planet
Occurrence (CPO) is helpful to understand the detailed shape of the
planet period distribution.  If planet occurrence is constant per
$\log \Per$ interval, CPO is a linear function in $\log \Per$. The
slope of the CPO conveys planet occurrence: the higher the planet
occurrence, the steeper the slope of the CPO.

\begin{figure*}
\centering
\includegraphics[width=.6\textwidth]{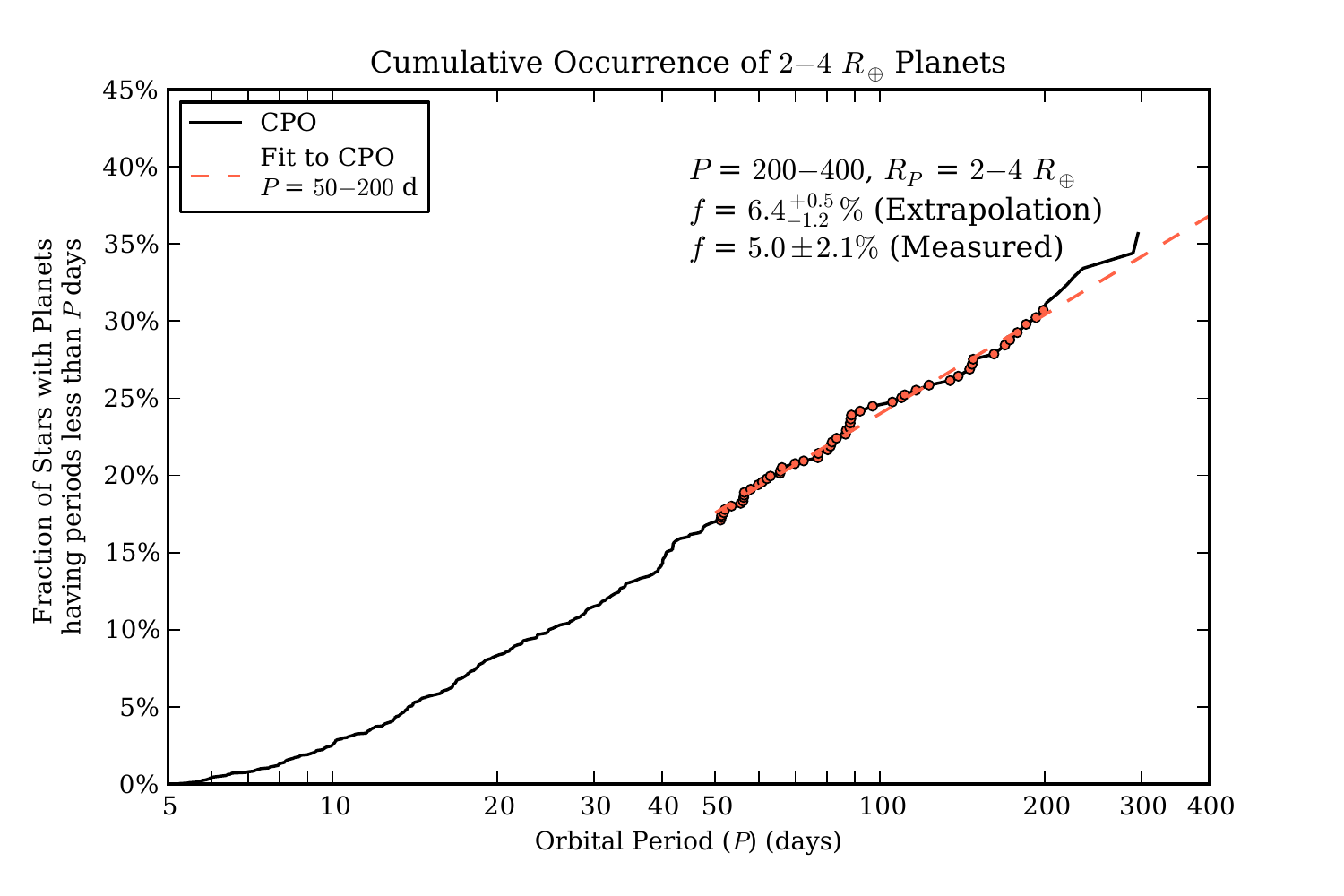}
\caption{The fraction of stars having 2--4~\Re planets with any
  orbital period up to a maximum period, \Per, on the horizontal
  axis. This is the Cumulative Planet Occurrence (CPO). A linear
  increase in CPO corresponds to planet occurrence that is constant in
  equal intervals of $\log \Per$.  The CPO steepens from 5 to
  $\sim10$~days, corresponding to increasing planet occurrence in the
  5--10~day range. For $\Per~\gtrsim~10$~days, the CPO has a constant
  slope, reflecting uniform planet occurrence per $\log \Per$
  interval. Planet occurrence in the \Rp~=~2--4~\Re,
  \Per~=~200--400~day domain is predicted to be \fBigYearExtrap by
  extrapolation from shorter periods, which is consistent with our
  measured value of \fBigYearMeas.}
\label{fig:CPO24}
\end{figure*}

Figure~\ref{fig:CPO24} shows CPO for \Rp~=~2--4~\Re planets. Planet
occurrence increases with period from 5 days up to $\sim10$~days, and
is consistent with uniform for larger periods. This change in the
planet period distribution was noted in previous
work~\cite{Youdin11,Howard12,Dong12}. We fit a line to the CPO from
50--200~days and extrapolate into the 200--400~day range. The
extrapolation predicts \fBigYearExtrap occurrence, which agrees with
our measured value of \fBigYearMeas to 1~$\sigma$. We estimate errors
on our extrapolation by fitting subsets of the CPO that span half the
original period range.  We fit 100 subsections ranging from
\Per~=~50--100~days up to \Per~=~100--200~days.

\begin{figure*}
\centering
\includegraphics[width=.6\textwidth]{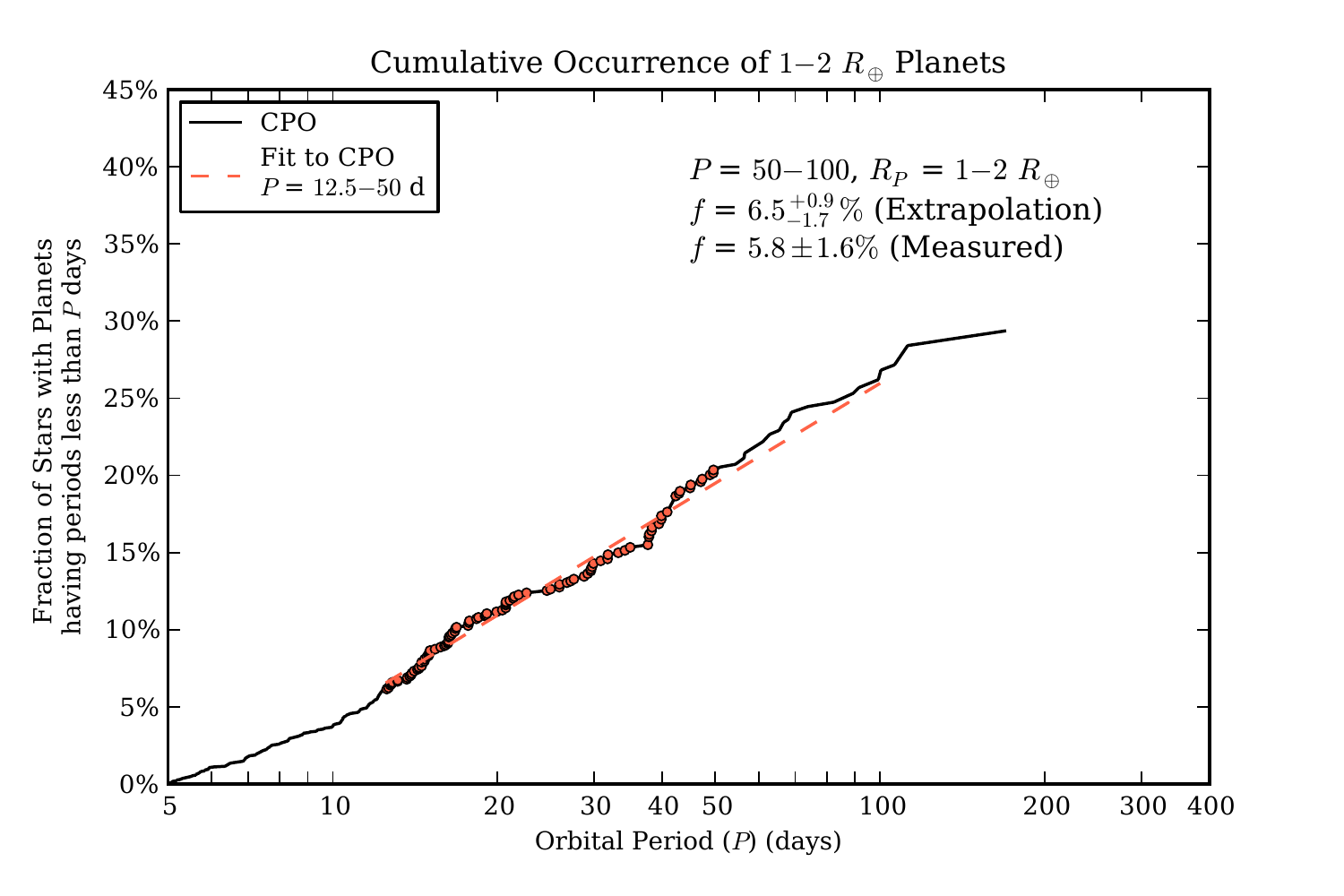}
\caption{Same as Figure~\ref{fig:CPO24}, but showing the CPO for 1--2
  \Re planets. Planet occurrence in the \Rp~=~1--2~\Re,
  \Per~=~50--100~day bin is predicted to be \fSmMercExtrap by
  extrapolation from shorter periods, which is consistent with our
  measured value of \fSmMercMeas.}
\label{fig:CPO12}
\end{figure*}

We also compare occurrence in the \Per~=~50--100~day, \Rp~=~1--2~\Re
domain based on extrapolation to our measured value. Figure~\ref{fig:CPO12}
shows the CPO for \Rp~=~1--2~\Re planets. We fit the CPO from
\Per~=~12.5--50~days. This fit predicts an occurrence of
\fSmMercExtrap in the 50--100~day range, in good agreement with our
measured value of \fSmMercMeas. The uniformity in the occurrence of
small planets as a function of period, lends support to the same kind
of modest extrapolation into the \Rp~=~1--2~\Re, \Per~=~200--400~day
domain.

\subsection{Planet Occurrence in the Habitable Zone}
We consider a planet to reside in the habitable zone if it receives a
similar amount of light flux, \Fp, from its host star as does the
Earth. As described in the main text, we consider the most recent
theoretical work on habitability of planets following the seminal work
by Kasting \cite{Kasting1993, Seager13, Kopparapu2013, Zsom2013,
  Pierrehumbert2011}.

We adopt an inner edge of the HZ
at 0.5 AU for a Sun-like star where a planet would receive four times
the light flux that Earth does. This inner edge is slightly more
conservative than that found by Zsom et al.~\cite{Zsom2013}.
The outer edge of the HZ less well
understood.  Kasting found the outer edge to be at 1.7~AU
\cite{Kasting1993}; Pierrehumbert and Gaidos \cite{Pierrehumbert2011}
found it could extend to 10~AU for planets with thick H$_2$
atmospheres. Here, we adopt an intermediate value of 2 AU for solar
analogs where the stellar flux is 1/4 that incident on the Earth. This
outer edge is consistent with the presence of liquid water on Mars in
its past. Mars might still have liquid water today, if it were more
massive.  Thus following the theory of planetary habitability, we
adopt a habitable zone for stars in general based on stellar flux
between 4x and 1/4 the solar flux falling on the Earth:
\Fp~=~0.25--4~\FE.

The stellar light flux hitting a planet, \Fp, depends linearly on
stellar luminosity, \Lstar, and inversely as the square of the
distance between the planet and the star.  Stellar luminosity, \Lstar,
is given by:
\[
\Lstar = 4\pi\Rstar^{2}\sigma\teff^{4},
\]
where $\sigma = 5.670 \times 10^{-8}$~W~m$^{-2}$~K$^{-4}$ is the
Stefan-Boltzmann constant.  In our study, the stellar radii and
temperatures, \teff, are computed two ways.  We obtained high SNR
spectra with high spectral resolution using the Keck Observatory HIRES
spectrometer for all of the 62 stars that host planets with periods
over 100 days, approaching the HZ.  For those 62 stars, we performed a
{\tt SpecMatch} analysis \cite{Petigura13} to determine \teff and the
surface gravity, \logg, and metalicity, [Fe/H].  These stellar values
were matched to stellar evolution models (Yonsei-Yale) to yield the
radii and masses of the stars.  The resulting values of stellar radii
are uncertain by 10\%, as determined by calibrations with nearby stars
having parallaxes and hence having more accurately determined stellar
radii.  The values of \teff are accurate to within 2\%.  Thus, summing
the fractional errors in quadrature, the resulting stellar
luminosities for the 62 stars (having \Per~>~100~days) are measured but
carry uncertainties of 25\%. For those stars without Keck spectra, 
we adopted photometric stellar radius and mass, for which the stellar radii
are in error by 35\% and the \teff values are 
uncertain by 4\%, giving errors in luminosity of 80\%.
We estimated the star-planet separation ($a$) using \Per, $\Mstar$,
and Kepler's third law.  The stellar light flux falling on a planet is
now easily calculated from \Fp $\propto$ \Lstar $/a^2$.  In what
follows, we quote the flux falling on a planet relative to that falling on the
Earth.

We find 10 planets having radii 1--2 \Re that fall within the stellar
incident flux domain of the habitable zone, 0.25--4 \FE. As a
reference, we plot their phase folded light curves in
Figure~\ref{fig:HZ10} along with the KIC identifier, period, radius,
and stellar light flux. To compute the prevalence of such planets
within the HZ, we apply the usual geometric correction for orbital
tilts too large to cause transits, augmenting the counting of each
transiting planet by $a/\Rstar$ total planets. We compute \Fp for each
synthetic planet in our completeness measurement
study. Figure~\ref{fig:HZComp} shows stellar flux level and radii of
the 10 habitable zone planets, having size 1--2~\Re, along with the
synthetic HZ planets from our completeness study. Because the
number of synthetic trials is small for \Fp~<~1~\FE, we compute
occurrence using the 8 planets with \Fp~=~1--4 \FE. We find $11\pm4\%$
of Sun-like stars have a \Rp~=~1--2~\Re planet that receives
\Fp~=~1--4~\FE light energy from their host star.

\begin{figure*}
\centering
\includegraphics[width=.6\textwidth]{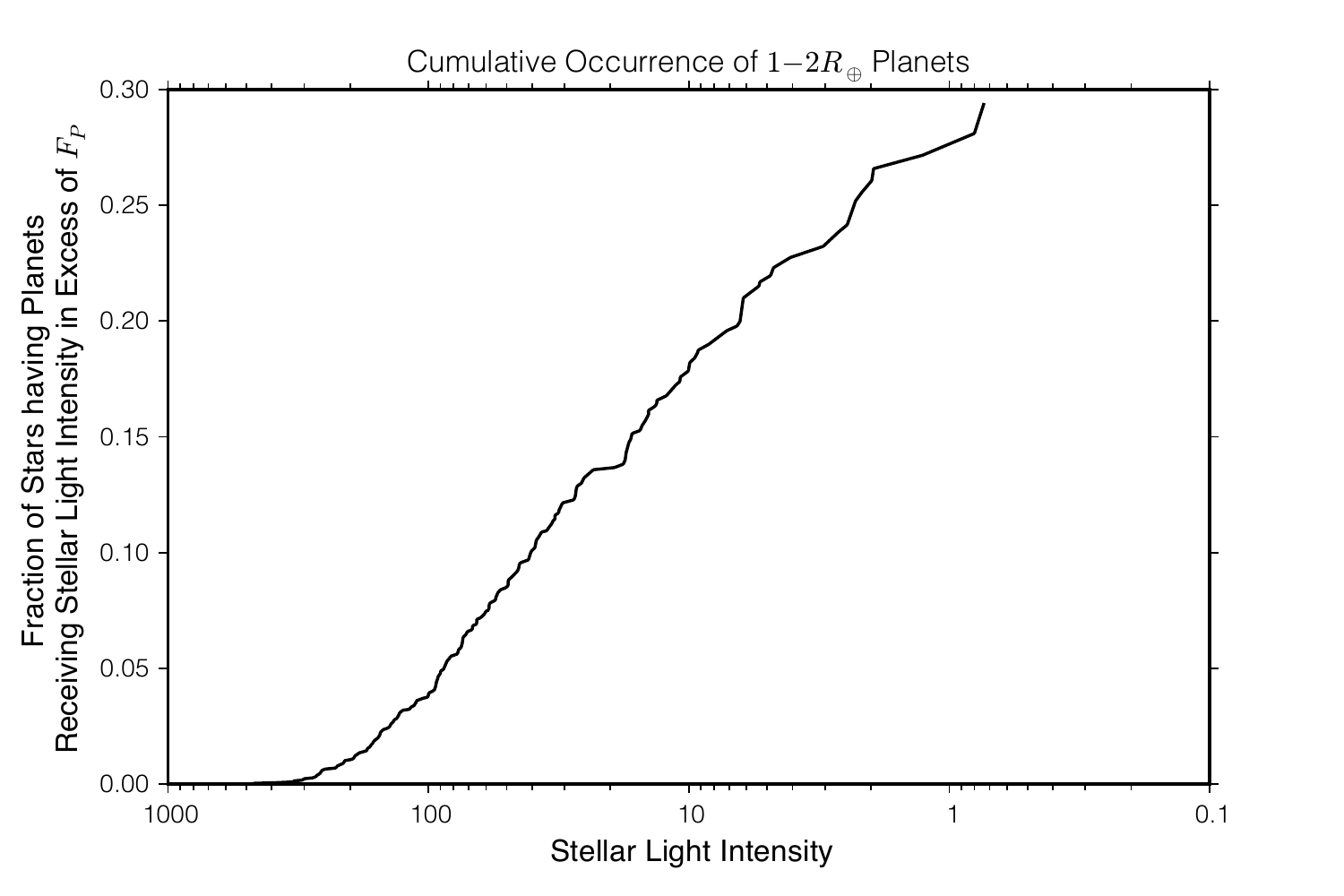}
\caption{Same as Figure~\ref{fig:CPO24}, but showing the CPO for 1--2
  \Re planets as a function of decreasing flux. Planet occurrence is
  constant over a wide range of incident flux values,
  \Fp~=~100--4~\FE, which supports extrapolation to the outer edge of
  our adopted HZ, \Fp~=~0.25~\FE.}
\label{fig:CPOFp}
\end{figure*}

\begin{figure}
\centering
\includegraphics[width=1\textwidth]{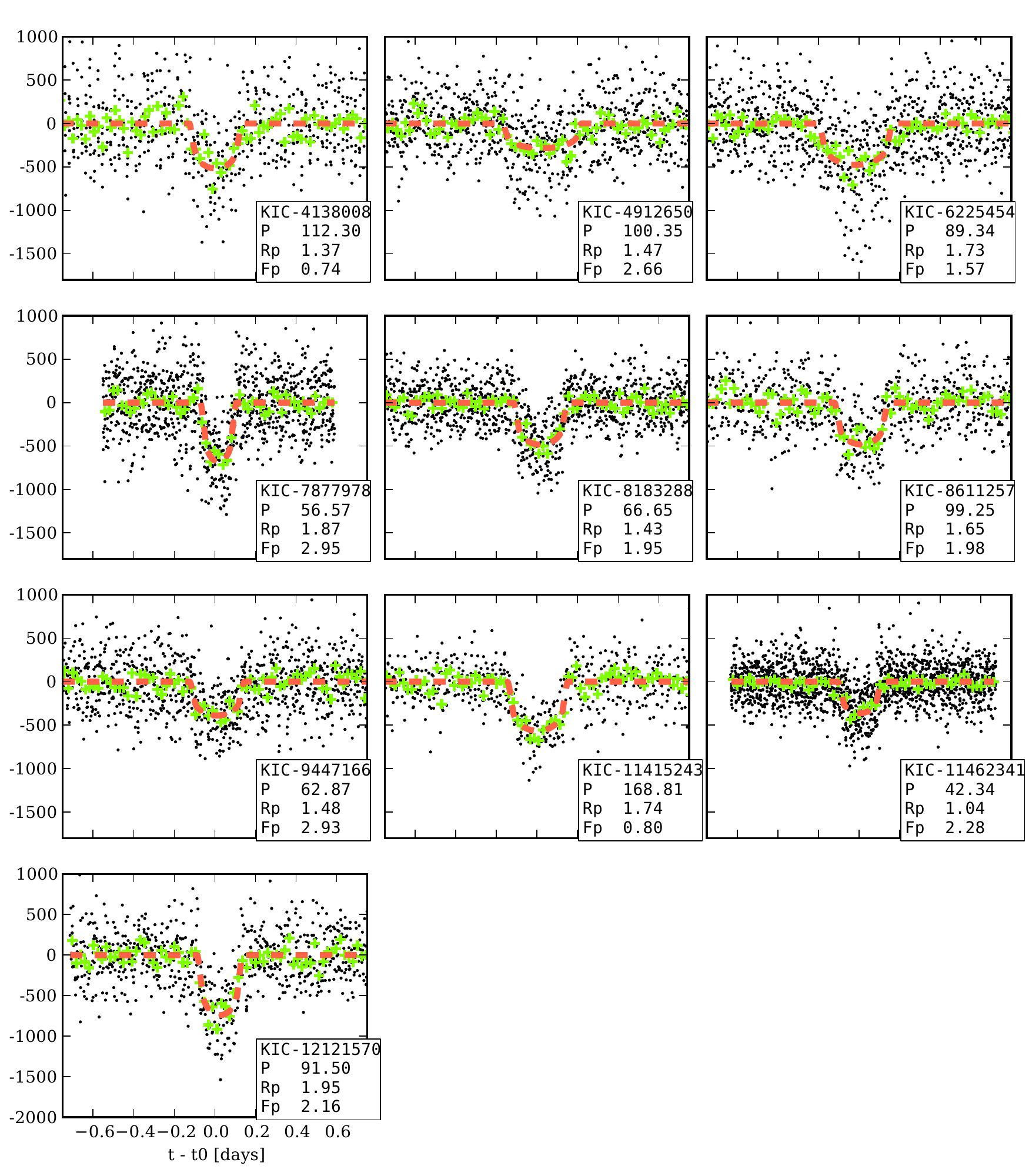}
\caption{Phase folded photometry for ten Earth-size HZ candidates. Black point
  shows show \TERRA-calibrated photometry folded on the best fit
  ephemeris listed in Table~\ref{tab:eKOI}. The green symbols show the
  median flux value in 30-minute bins. The red dashed lines shows the
  best-fit Mandel-Agol model. We have annotated each plot with the KIC
  identifier, period, planet size (Earth-radii), incident flux level
  (relative to Earth). All measurements of planet size and incident flux
  are based on spectra taken with the Keck 10~m telescope. Spectra for 
  KIC-6225454, KIC-7877978, KIC-9447166, and KIC-11462341 were obtained during
  peer-review and were added in proof (see Table~\ref{tab:eKOIinpress}).}
\label{fig:HZ10}
\end{figure}

\begin{figure*}
\includegraphics[width=1\textwidth]{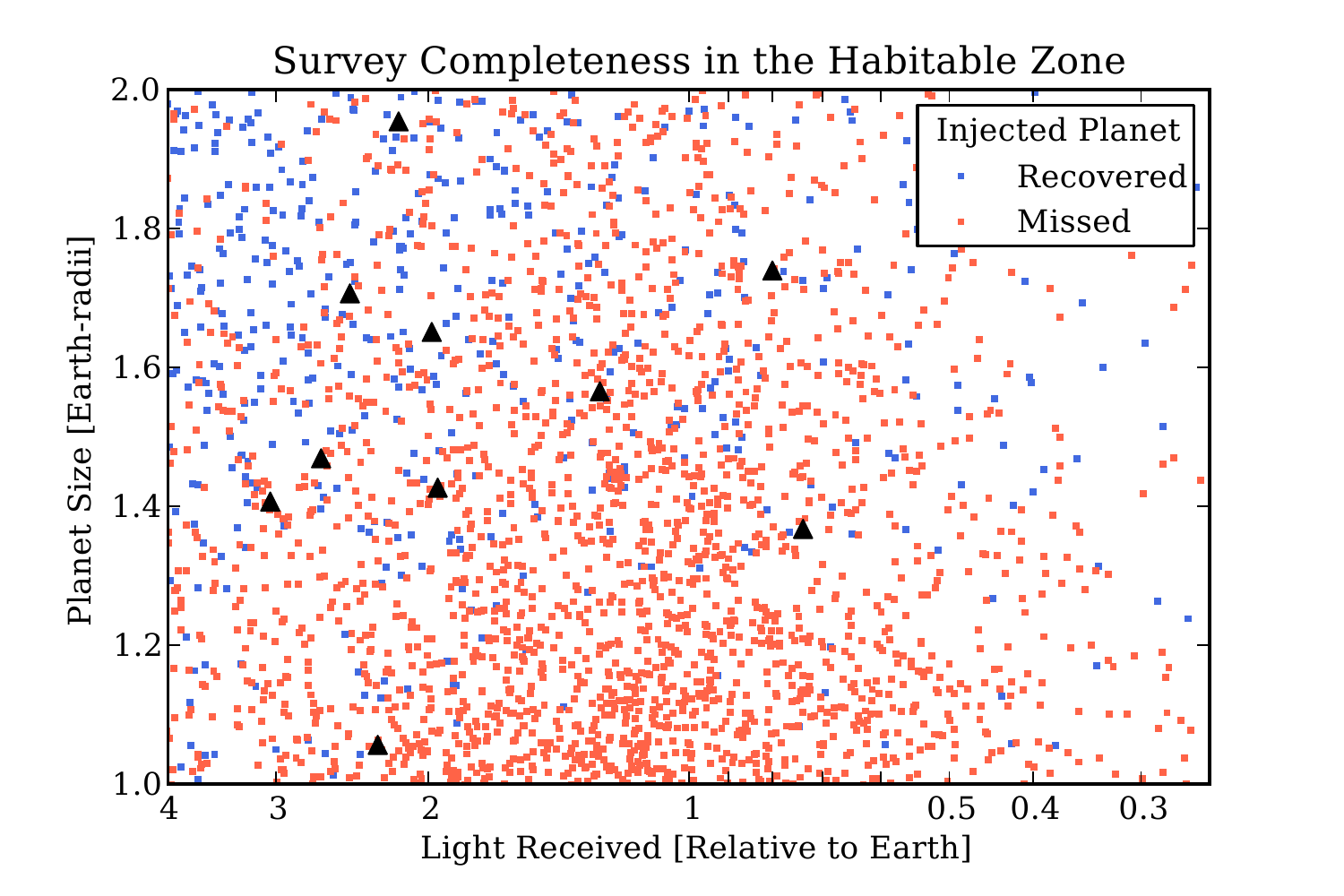}
\caption{Ten small (\Rp~=~1--2~\Re) planets (black triangles) fall
  within our adopted habitable zone of \Fp~=1/4--4~\FE. We also plot
  the injected planets over this same domain. Survey completeness $C$
  is computed locally for each planet by dividing the number of
  injected planets that were recovered by the total number of injected
  planets in a small box centered on the real planet.}
\label{fig:HZComp}
\end{figure*}

We account for the entire HZ (extending out to 0.25~\FE) by
extrapolating occurrence in \Fp, assuming constant planet occurrence
per $\log \Per$ interval. Figure~\ref{fig:CPOFp} shows the CPO as a
function of \Fp. Planet occurrence is constant from $\sim100~\FE$ down
to $\sim4~\FE$, beyond which, small number fluctuations are
significant. Assuming the occurrence of planets is constant in $\log
\Fp$ implies that the same number of 1--2 \Re planets have incident
fluxes of 1--0.25 \FE as have fluxes of 1--4 \FE where we computed
directly the occurrence of planets to be 11\%.  Thus, $22\pm8\%$ of
Sun-like stars have a \Rp~=~1--2~\Re planet within our adopted
habitable zone with fluxes of 0.25-4.0 \FE.

\clearpage

\subsection{Occurrence Including Planets in Multi-planet Systems}
For systems harboring more than one planet, \TERRA only detects the
planet with highest SNR, i.e. the most significant planet. The actual
rate of planet occurrence is higher than we report when the missed
planets in these multi-transiting systems is included.  (Note that
this correction only applies to multi-transiting system and not all
multi-planet systems.) We estimate the size of this effect using the
Q12 sample of KOIs from the \Kepler project, which includes stars with
multiple planets. We selected the 1190 ``candidates'' with
well-determined periods ($\sigma(\Per) < 0.1$~days) that orbit stars in
the Best42k. In order to make a fair comparison between our planet
sample and the Q12 sample, we computed the SNR of each of the 1190
candidates using \TERRA. We excluded 82 KOIs with SNR < 12,
i.e. candidates that would have been deemed sub-significant by \TERRA.

\begin{figure}
\centering
\includegraphics[width=0.8\textwidth]{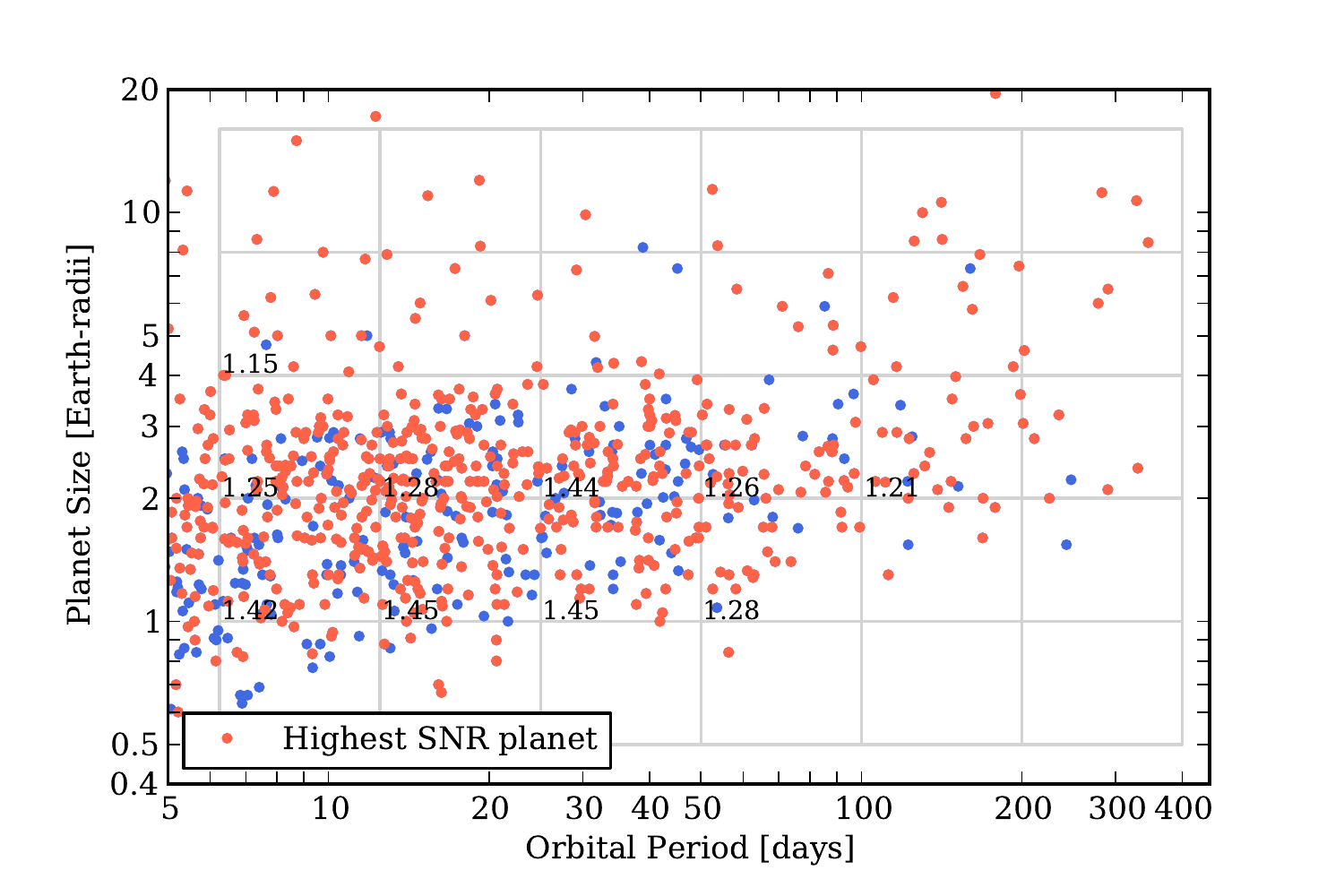}
\caption{-- Distribution of \Per and \Rp for \Kepler team candidates
  from the Q12 catalog. We include planets with \TERRA SNR > 12 and
  well-determined orbital periods ($\sigma(\Per) < 0.1$~days) that around
  ``Best42k'' stars. Planets that are either single or are the most
  significant planet in a multi-planet system are shown in red. Blue
  points correspond to additional planets in multi planet systems. For
  each cell with 10 or more planets, we compute the boost in planet
  counts due to multiplanet systems, the total number of planets
  divide by the number of most significant planets. For planets
  smaller than 4~\Re and \Per~50~days, the boost ranges from
  21--28\%. Thus, including multis, we expect $\sim25\%$ higher
  occurrence.}
\label{fig:MultiBoost}
\end{figure}

For each planet in a multi-transiting system, we rank order each
candidate by its ``Relative SNR'' defined as:
\[
\text{Relative SNR} = \df \sqrt{ \frac{\tdur}{\Per} }.
\]
Figure~\ref{fig:MultiBoost} shows the distribution of Q12 candidates in
the Best42k as points on the \Per--\Rp plane. We highlight points
corresponding to the most significant planet. We assess the boost in
planet counts due to multi-transiting systems for different domains in
\Per and \Rp. For \Per~>~50~days and \Rp~<~4~\Re, this multi-boost factor
ranges from 21 to 28\%, neglecting bins with fewer than 8 detected
planets that suffer from small number fluctuations. Had we included
additional planets, our occurrence measurements would rise by
$\sim25\%$, which is comparable to or slightly smaller than the
fractional occurrence error for small planets in long-period orbits.

\subsection{Correction due to False Positives}
As discussed earlier, the sample of eKOIs is polluted by astrophysical
false positives. Like the \Kepler team, we do our best to identify and
remove transits that are clearly due to eclipsing binaries, but cannot
remove all eclipsing binary configurations. Thus, our sample, as well
as those produced by the Kepler team, still contain a false positive
component.

Fressin et al. (2013) addressed the contamination of the February 2012
\Kepler Project sample of KOIs~\cite{Batalha12} by FPs that were not
removed by the \Kepler Project vetting process. FPs include background
eclipsing binaries, physically associated eclipsing binaries
(hierarchical triples), and physically associated stars, which
themselves have a transiting planet. We consider the last scenario to
be a FP because even though the transiting object is a planet, the
radius is at least 1.4 times larger (1.4 corresponds stars of equal
brightness). Fressin et al. (2013) added FPs to the Batalha et al
(2012) sample of KOIs according to models of galactic structure,
stellar binarity, and assumptions about the distributions of
planets. Simulated FPs that would exhibit a detectable secondary
eclipse or a significant centroid offset were removed, assuming the
Kepler Project vetting process catches these FPs.

Fressin et al. (2013) found an overall FP rate of $8.8\pm1.9\%$
for 1.25--2.0~\Re planets and $12.3\pm3.0\%$ for 0.8--1.25~\Re
planets. Again, note that this fractional occurrence correction is
small compared to our reported errors for small planets in long-period
orbits. Stars with bound companions with transiting planets are the
dominant fraction of FPs for small planets (76\% for 1.25--2.0~\Re planets and
66\% for 0.8--1.25~\Re planets). FPs of this type are very difficult
to identify. A Sun-like star with $V$~=~14.7~mag (typical for our
sample) is 1~kpc away. The binary star separation distribution peaks at
50 AU~\cite{Raghavan10} or 0.05 arcsec assuming a face-on
orbit. Detecting companions separated by 0.05 arcsec is near the
limits of current ground-based AO. Even if a companion was detected,
we still wouldn't know which star harbored the transiting planet.

We adopt a 10\% FP rate for planets having \Per~=~50--400~days and
\Rp~=~1--2~\Re. Adopting a false positive rate that is constant with
period is justified because the occurrence of Neptune-sized planets is
approximately constant with period, as shown in the main text. In the
context of the occurrence of Earth-size planets with
\Per~=~200--400~days (\EtaEarthErr) and Earth-size planets in the HZ
($22\pm8\%$), FPs contribute 10\% fractional uncertainty and are
secondary compared to statistical uncertainty.

\clearpage
\bibliographystyle{pnas}
\bibliography{terra1yr_si}
\clearpage

\begin{deluxetable}{rrrrlrrrrrrlrrrr}
\tablenum{S2}
\tabletypesize{\scriptsize}
\setlength{\tabcolsep}{0.05in}
\tablecaption{Properties of \neKOI eKOIs}
\tablewidth{0pt}
\tablehead{
  \colhead{KIC} &
  \colhead{KOI} &
  \colhead{\Per} &
  \colhead{\ep} &
  \colhead{Disp.} &
  \colhead{$p$} &
  \colhead{$\sigma(p)$} &
  \colhead{\teff} &
  \colhead{\logg} &
  \colhead{\Rstar} &
  \colhead{$\sigma(\Rstar)$} &
  \colhead{Prov.} &
  \colhead{\Rp} &
  \colhead{$\sigma(\Rp)$} &
  \colhead{\Fp}\\
  \colhead{} &
  \colhead{} &
  \colhead{days} &
  \colhead{days} &
  \colhead{} &
  \colhead{\%} &
  \colhead{} &
  \colhead{K} &
  \colhead{cgs} &
  \colhead{\Rsun} &
  \colhead{} &
  \colhead{} &
  \colhead{\Re} &
  \colhead{} &
  \colhead{\FE} 
}
\startdata
  1026032 &          &   8.460 & 133.78 &    R & 40.6 &  0.09 & 6044 &   4.5 &  0.95 &   0.29 &    P &  42.12 & 12.64 & 159.7 \\
  1026957 &   958.01 &  21.762 & 144.77 &    P &  3.1 &  0.42 & 4861 &   4.6 &  0.72 &   0.07 &   SM &   2.46 &  0.41 &  13.4 \\
  1571511 &   362.01 &  14.023 & 135.52 &   SE & 12.8 &  0.00 & 6055 &   4.4 &  1.04 &   0.31 &    P &  14.46 &  4.34 &  98.7 \\
  1718189 &   993.01 &  21.854 & 144.32 &    P &  1.7 &  0.10 & 5827 &   4.5 &  0.88 &   0.26 &    P &   1.65 &  0.50 &  34.6 \\
  1849702 &  2538.01 &  39.833 & 155.62 &    P &  1.7 &  0.37 & 5023 &   4.6 &  0.73 &   0.07 &   SM &   1.37 &  0.33 &   7.1 \\
  1872821 &  2351.01 &  10.274 & 134.64 &    P &  1.8 &  0.11 & 5706 &   4.6 &  0.85 &   0.25 &    P &   1.69 &  0.52 &  82.7 \\
  2141783 &  2201.01 & 116.521 & 144.14 &    P &  2.5 &  0.37 & 6063 &   4.2 &  1.53 &   0.15 &   SM &   4.09 &  0.74 &  10.9 \\
  2164169 &  1029.01 &  32.312 & 133.88 &    P &  2.3 &  0.07 & 5975 &   4.5 &  0.96 &   0.29 &    P &   2.39 &  0.72 &  26.5 \\
\enddata
\label{tab:eKOI}
\tablecomments{For each of the \neKOI eKOIs, we list the target star
  identifier, ephemeris, false positive status, transit fit
  parameters, host star properties, and planet radius. KIC --- \Kepler
  Input Catalog~\cite{Brown11} identifier. KOI --- \Kepler team
  identifier, if eKOI appears in 13 September 2013 cumulative list of
  candidates from the NASA Exoplanet
  Archive~\cite{Akeson13}. Disp. --- disposition according to the
  false positive vetting described in~\ref{sec:AFP}. eKOIs may be
  designated as a false positives for any of the following reasons:
  `SE' --- secondary eclipse, `VD' --- variable depth transits, `TTV'
  --- large transit timing variations, `V' --- V-shaped transit, `C'
  --- centroid offset. If an eKOI passes all the vetting steps, it is
  considered a planet, `P.' $p$ --- planet to star radius ratio,
  \rrat. Prov. --- provenience of stellar parameters: SpecMatch `SM'
  or photometric `P'.}
\end{deluxetable}

\begin{deluxetable}{rrrrlrrrrrrlrrrr}
\tabletypesize{\scriptsize}
\tablenum{S3}
\setlength{\tabcolsep}{0.05in}
\tablecaption{Spectroscopic properties of 13 eKOIs (added in proof)}
\tablewidth{0pt}
\tablehead{
  \colhead{KIC} &
  \colhead{KOI} &
  \colhead{\Per} &
  \colhead{\ep} &
  \colhead{Disp.} &
  \colhead{$p$} &
  \colhead{$\sigma(p)$} &
  \colhead{\teff} &
  \colhead{\logg} &
  \colhead{\Rstar} &
  \colhead{$\sigma(\Rstar)$} &
  \colhead{Prov.} &
  \colhead{\Rp} &
  \colhead{$\sigma(\Rp)$} &
  \colhead{\Fp}\\
  \colhead{} &
  \colhead{} &
  \colhead{days} &
  \colhead{days} &
  \colhead{} &
  \colhead{\%} &
  \colhead{} &
  \colhead{K} &
  \colhead{cgs} &
  \colhead{\Rsun} &
  \colhead{} &
  \colhead{} &
  \colhead{\Re} &
  \colhead{} &
  \colhead{\FE} 
}
\startdata
  4478142 &          & 219.909 & 329.43 &    P &  2.5 & 0.34 & 5648 &   4.3 &  1.06 &   0.11 &   SM &  2.92 &  0.49 & 2.3 \\
  4820550 &  3823.01 & 202.121 & 292.04 &    P &  5.7 & 0.65 & 5594 &   4.5 &  0.94 &   0.09 &   SM &  5.79 &  0.88 & 1.8 \\
  6225454 &          &  89.338 & 218.54 &    P &  2.4 & 0.58 & 4652 &   4.6 &  0.68 &   0.07 &   SM &  1.73 &  0.46 & 1.6 \\
  6307083 &  2050.01 &  75.378 & 167.11 &   SE &  1.7 & 0.09 & 5109 &   4.6 &  0.77 &   0.08 &   SM &  1.43 &  0.16 & 3.4 \\
  7101828 &   455.01 &  47.878 & 145.35 &   VD &  2.2 & 0.10 & 4328 &   4.7 &  0.55 &   0.05 &   SM &  1.33 &  0.15 & 2.0 \\
  7866914 &  3971.01 & 366.020 & 182.66 &   SE &  4.0 & 1.15 & 5713 &   4.3 &  1.06 &   0.11 &   SM &  4.67 &  1.40 & 1.2 \\
  7877978 &  2760.01 &  56.573 & 146.59 &    P &  2.5 & 0.35 & 4675 &   4.6 &  0.68 &   0.07 &   SM &  1.87 &  0.32 & 3.0 \\
  8044608 &  3523.01 & 106.176 & 234.62 &    R & 55.2 & 2.76 & 6056 &   4.4 &  1.12 &   0.11 &   SM & 67.58 &  7.56 & 7.7 \\
\enddata
\label{tab:eKOIinpress}
\tablecomments{We obtained 13 additional spectra of long period eKOIs during peer-review. In this work, we used photometric properties for these 13 eKOIs, but we include them here for completeness. Column descriptions are the same as Table~\ref{tab:eKOI}.}
\end{deluxetable}
\clearpage
\end{document}